\documentclass{aa}
\usepackage{graphicx}
\usepackage{txfonts}

\usepackage{graphicx}
\usepackage{txfonts}
\usepackage{graphicx}
\usepackage{lscape}
\usepackage{needspace}
\usepackage{placeins}
\usepackage{supertabular}
\usepackage{caption}
\usepackage[colorlinks=true, linkcolor=blue, citecolor=blue, filecolor=blue, urlcolor=blue]{hyperref}
\usepackage[version=3]{mhchem}
\usepackage{xcolor}
\usepackage{lscape}
\usepackage{ulem}
\usepackage{mhchem}
\usepackage{longtable}
\usepackage{caption}
\usepackage{booktabs}
\usepackage{hyperref}
\usepackage{array}
\usepackage[figuresright]{rotating}
\usepackage{mathtools}
\usepackage{float}
\usepackage{csquotes}

\begin{document}

\title{Probing sulfur chemistry in oxygen-rich AGB stars with ALMA}

   \author{P. Gorai
          \inst{1,2}\fnmsep\thanks{prasanta.astro@gmail.com}
          \and
         M. Saberi\inst{1,2}
         \and
         T. Khouri\inst{3}
         \and
         T. Danilovich\inst{4,5}}

   \institute{Rosseland Centre for Solar Physics, University of Oslo, PO Box 1029 Blindern, 0315, Oslo, Norway
              \and
              Institute of Theoretical Astrophysics, University of Oslo, PO Box 1029 Blindern, 0315, Oslo, Norway
              \email{prasanta.astro@gmail.com}
              \and
              Dep. of Space, Earth and Environment, Chalmers University of Technology, Onsala Space Observatory, 43992 Onsala, Sweden
              \and
              School of Physics \& Astronomy, Monash University, Wellington Road, Clayton 3800, Victoria, Australia
              \and Institute of Astronomy, KU Leuven, Celestijnenlaan 200D,  3001 Leuven, Belgium }

\date{Oct, 2025; Accepted}
  \abstract  
   {Sulfur and its isotopic ratios play a crucial role in understanding the physical properties of astrophysical environments, particularly providing key insights into nucleosynthesis, interstellar medium (ISM) processes, star formation, planetary system evolution, and galactic chemical evolution.}
   {We aim to investigate the distribution of sulfur species- SO$_2$, $\rm{^{34}SO_2}$, SO, and $\rm{^{34}SO}$—toward a sample of five oxygen-rich Asymptotic Giant Branch (AGB) stars, along with measurements of excitation temperature, column density, and isotopic ratios.}
   {We use ALMA Band 6, 7, and 8 data of: {\it o} Ceti, R Dor, W Hya, R Leo, and EP Aqr. SO$_2$, $\rm{^{34}SO_2}$, SO, and $\rm{^{34}SO}$ are detected towards AGB stars using the CASSIS software. To estimate the gas temperature and column density of these species, we apply the rotational diagram method (when applicable) and the Markov Chain Monte Carlo (MCMC) method, assuming Local Thermodynamic Equilibrium (LTE). Finally, line imaging of different transitions is performed to infer the distributions of the detected sulfur-bearing species in our sample.}
   {The measured excitation temperatures of SO$_2$ for our sample sources range from $\sim$ 200 to 600 K, with estimated column densities in the range of $\rm{1-7\times10^{16}\ cm^{-2}}$. The excitation temperatures estimated using $\rm{^{34}SO_2}$ are comparable or slightly lower, while the column densities are about an order of magnitude lower than those of SO$_2$. 
   Our measured $\rm{^{32}S}$/$\rm{^{34}S}$ ratio for R Dor and W Hya are close to the solar value, however, the measured value for {\it o} Ceti is slightly higher, and the measured values for EP Aqr and R Leo are lower. Finally, spatial analysis shows that most detected lines appear as centralized emissions. Moreover, the high excitation transitions of SO$_2$ show compact emission and probe hot gas of the inner region circumstellar envelopes (CSEs), whereas low-excitation transitions trace slightly extended structures. However, we find some differences in the emission of detected species across our sample.}
   {The excitation temperature of the observed regions of the CSE can be probed using the SO$_2$ molecule. The morphological correlation between SO and SO$_2$ emissions suggests they are chemically linked. Differences in the emission distributions of the detected species across our sample of low mass-loss rate AGB stars such as (i) centralized emission towards {\it o} Ceti with irregular emission shapes, (ii) centralized emission with ordered circular features towards R Leo and W Hya, (iii) clumpy emission features in R Dor, and (iv) unresolved emission in Ep Aqr may arise from several factors, i.e., the physical conditions of the sources (e.g., density and temperature structures of the CSEs), source multiplicity, outflows, rotation, or other associated physical processes such as thermal and non-thermal desorption, the effects of UV photons and cosmic rays, and finally the resolution of our observations. Nonetheless, the predominantly centralized distributions of SO and SO$_2$ in our sample support previous findings for low mass-loss rate AGB stars. Our measured $\rm{^{32}S}$/$\rm{^{34}S}$ ratios for the two stars R Dor and W Hya agree well with solar values within uncertainties, indicating that these ratios likely reflect the isotopic composition of the stars' natal clouds and deviate for three stars ({\it o} Ceti, R Leo and EP Aqr) which could be due to the metallicity and/or excitation conditions within various sources.}


   \maketitle
%

\section{Introduction}

When a star with an initial mass of 1-8 M$_\odot$ approaches the end of its life and enters the asymptotic giant branch (AGB) phase, during which it rapidly loses mass (in the range $\rm{10^{-8} - 10^{-4}\ M_{\odot}\ yr^{-1}}$) and consequently creates an expanding circumstellar envelope (CSE). 
In the AGB phase, the star undergoes periodic events known as thermal pulses, which trigger dredge-up episodes that bring carbon from the interior to the surface. If the carbon abundance at the surface exceeds that of oxygen, the star transforms into a carbon star.  Depending on the C/O ratio, an AGB star can be classified as oxygen-rich when C/O $<$ 1, carbon-rich when C/O $>$ 1, and S-type where C/O $\sim$ 1.

Sulfur is the tenth most abundant element by mass in the Universe, and it plays a crucial role in biological systems \citep{Mifsud2021,Krijt2023}. Sulfur has four stable isotopes with fractional abundances of $^{32}$S (94.85\%), $^{33}$S (0.76\%), $^{34}$S (4.36\%), and $^{36}$S (0.02\%), reflecting their relative abundances at the time of the birth of the Sun \citep[][]{Asplund21}. The sulfur isotope ratio provides crucial and complementary information on stellar synthesis that is not traced by carbon \citep{Yan2023}. 
It is believed that $\rm{^{32}S}$ and $\rm{^{34}S}$ are mainly synthesized during the oxygen-burning process of Type II and Type Ia supernovae (SNe) and $\rm{^{33}S}$ synthesized in explosive oxygen and neon-burning \citep{Woosley95,Yan2023}.

\cite{Asplund21} reported a ratio of $\rm{^{32}S}$/$\rm{^{34}S} \sim 21.7$ for the sun. \cite{Danilovich2020} estimated $\rm{^{32}S/^{34}S}$ ratios for two oxygen-rich AGB stars, 18.5$\pm$5.8 for R Dor in agreement with the solar value, and 42 for IK Tau, which is considerably higher than the solar value. Recently, \cite{Wallstrom2024} measured the $\rm{^{32}S/^{34}S}$ ratio for the S-type star, W Aql as 16.7$\pm$5.6, and two high-mass loss rate oxygen-rich AGB stars, IRC+10011 and IRC-10529 as 25$\pm$12.5, and 9.10$\pm$5, respectively. Their measured $\rm{^{32}S/^{34}S}$ ratios are consistent with the solar value within uncertainties for W Aql and IRC+10011, but differ for IRC-10529 star. \cite{Unnikrishnan2024} reported this ratio for three carbon-rich AGB stars, 15194-5115, 15082-4808, and 07454-7112 as 19$\pm$4, 28$\pm$8, and 21$\pm$4, respectively. These measured values are in good agreement with the solar value, considering their uncertainties reported, except for the high-mass oxygen-rich AGB star, IRC-10529. The reported lower value for IRC-10529 compared to the solar value of 21.7 is not discussed. We speculate that comparing the measured ratio values with the solar value may highlight variations in elemental abundances between the natal clouds of AGB stars and those of the Sun.

Several sulfur-bearing species, such as CS, SO, SO$_2$, and H$_2$S, are commonly detected in the circumstellar environments of AGB stars and star-forming regions \citep[e.g.,][]{Omont93, Danilovich2017,Massalkhi2020,Fontani2023,Ghosh2024}.
Several studies of sulfur towards oxygen-rich AGB stars have been carried out \citep[e.g.,][]{Lindqvist1988,Danilovich2016,Danilovich2017,Danilovich2018,Danilovich2020}. \cite{Danilovich2016} analyzed the SO and SO$_2$ spatial distributions using single dish observations toward five oxygen-rich AGB stars (IK Tau, R Dor, TX Cam, W Hya, R Cas) with different mass-loss rates. 

They found that higher mass-loss rate stars show shell-like distributions of SO and lower peak relative abundances. The lower mass-loss rate stars show centralised SO distributions with higher peak abundances close to the stars. The locations of the SO peaks (for the higher mass-loss rate stars) and the e-folding radii (for the lower mass-loss rate stars) were correlated with the OH peak abundance and the photodissociation of H$_2$O. Later, \cite{Danilovich2020} presented the spatially resolved observations of SO and SO$_2$ toward two AGB stars, one with a high-mass loss rate, IK Tau, and another with a low-mass loss rate, R Dor. In the case of R Dor, the emission of two sulfur species coincides with peaks around the central star, which trace out the same density structures in the circumstellar environment. In the case of IK Tau, SO shows a shell-like structure; its peak does not appear at the star's center, and most of the flux was resolved for the low excitation SO$_2$ transition. 

\cite{Massalkhi2020} observed SO and SO$_2$ in oxygen-rich CSE of AGB stars. They further compared their distribution with SiO to determine whether SO and SO$_2$ can be used as a dust precursor. They looked into the SiO abundance profile and found a decreasing trend with the envelope density, similar to C-rich AGB stars, indicating SiO adsorption onto the dust grains. Interestingly, they obtained a similar trend in the case of SO, but not as prominent as for SiO. On the other hand, SO$_2$ does not show such a trend, which indicates it is a less important candidate to be the precursor of the dust. Recently, \cite{Wallstrom2024} presented high-resolution ALMA observations of SO and SO$_2$ in a large sample of 17 oxygen-rich AGB stars with a variation of mass-loss rates. They show that the spatial distributions of SO and SO$_2$ are generally consistent with previous results, with a centralized distribution for low mass-loss rate sources and a shell-like distribution for high mass-loss rate sources.

In this paper, we report several transitions of SO, SO$_2$, and their isotopologues $\rm{^{34}SO}$ and $\rm{^{34}SO_2}$ in a wider energy range observed with ALMA Bands 6, 7, and 8 towards a sample of 5 oxygen-rich AGB stars, all of which have mass-loss rates on the order of $\sim$ 10$^{-7}$ $\rm{M_{\odot}\ yr^{-1}}$. The paper is organized as follows: Section \ref{sec:2} describes the observations and data analysis, and the Results are presented in Section \ref{sec:3}. Discussions are summarized in Section \ref{sec:4}, and finally, the conclusion is given in Section \ref{sec:5}.

\begin{table*}[h!]
\caption{Our sample: M-type AGB stars}
\label{tab:targets}
\centering
\begin{tabular}{cccccccccccc}
\hline
Star & RA &DEC& $\dot M$ & d & $V_{\rm LSR}$ & $V_{\rm exp}$ & $R_{\star}$ &  $T_{\star}$  & P & $M_\star$\\
    &(J2000)&(J2000) & \tiny($10^{-7} M_{\odot} yr^{-1}$) & \tiny(pc) & \tiny(km s$^{-1}$)& \tiny(km s$^{-1}$) & \tiny(mas) & \tiny(K)  & \tiny(days) &\tiny($M_{\odot}$)
     \\ 
\hline
{\it o} Ceti&02$^{\mathrm{h}}$19$^{\mathrm{m}}$20$^{\mathrm{s}}$.795&-02$^{\circ}$58$^{\prime}$47.313$^{\prime\prime}$
 &  2.0 & 102 & 47 & 3 & 15 &  2800 & 332 & 1.0 \\
R Leo&09$^{\mathrm{h}}$47$^{\mathrm{m}}$33$^{\mathrm{s}}$.480 &+11$^{\circ}$25$^{\prime}$42.883$^{\prime\prime}$
 &  1.0 & 130 & 0.5 & 6 & 13.5 &2800  & 310 & 1.5 \\ 
W Hya &  13$^{\mathrm{h}}$49$^{\mathrm{m}}$01$^{\mathrm{s}}$.925& -28$^{\circ}$22$^{\prime}$04.648$^{\prime\prime}$
 & 1.3 & 104 & 41 & 7 & 20 &  2950 & 388 & 1.0 \\ 
R Dor &04$^{\mathrm{h}}$36$^{\mathrm{m}}$45$^{\mathrm{s}}$.400  & -62$^{\circ}$04$^{\prime}$39.259$^{\prime\prime}$
  & 1.6 & 45 & 6.9 & 6 & 27.5 &  2400 & 175 & 1.0-1.3 \\ 
EP Aqr &  21$^{\mathrm{h}}$46$^{\mathrm{m}}$31$^{\mathrm{s}}$.880 &-02$^{\circ}$12$^{\prime}$45.547$^{\prime\prime}$
 &1.2 & 114 & -34 & 11 & -- &  --& -- & -- \\ 

\hline
\end{tabular}
\tablefoot{Columns 4 to 11 \citep[Same as reported in][]{saberi2022}, for the first four sources, are taken from \citet[][]{Vlemmings19}. The stellar masses for W Hya and R Dor are from \cite{Taissa2017}, while those for $o$ Ceti and R Leo are estimated based on the $^{17}$O/$^{18}$O ratio presented by \cite{DeNutte17}, and the O isotopic ratios are taken from \cite{Hinkle16}. For EP Aqr, all parameters are taken from \citet[][and references therein]{Nhung2015}. We note that $R_\star$ listed here are the measured radii in the near-infrared. \dag Period of R Dor varies between 175 and 362 days.} 
\end{table*}

\section{Observations and data analysis \label{sec:2}}
In this paper, we used the Atacama Large Millimeter/submillimeter Array (ALMA) Band 8 observations with the Atacama Compact Array (ACA) (2018.1.01440.S, PI: M. Saberi) configuration for four sources (R Dor, R Leo, W Hya, and EP Aqr)  and the 12m array (2018.1.00649.S, PI: M. Saberi) for one source ({\it o} Ceti). Additionally, we used 12m array observations of {\it o} Ceti with ALMA Band 7 (2018.1.00749.S, PI: T. Khouri). For R Dor, we used ALMA 12m array Band 7 observations (2017.A.00012.S, PI: L. Decin). We also used 12m array observations of {R Leo} with the ALMA Band 7 setup (2019.1.00801.S, PI: J. Champion) and 12m array observations of W Hya with ALMA Band 6 (2016.1.00374.S, PI: K. Ohnaka). Notably, we used a more or less similar angular resolution for all sources: approximately $\sim$2.5" with ACA and $\sim$0.2" with the 12m array. The targets, along with their coordinates, distances, LSR velocity (V$_{\rm LSR}$), expansion velocities, masses, temperatures, radii, and mass-loss rates, are listed in Table~\ref{tab:targets}. Table~\ref{tab:obs-summary} provides a summary of the observations, which includes information about ALMA Band, configuration, no. of antennas, observation date, angular resolution, maximum recoverable scale (MRS), field of view (FOV), sensitivity, and spectral resolution.

\begin{figure*}
    \centering
    \includegraphics[width=\textwidth]{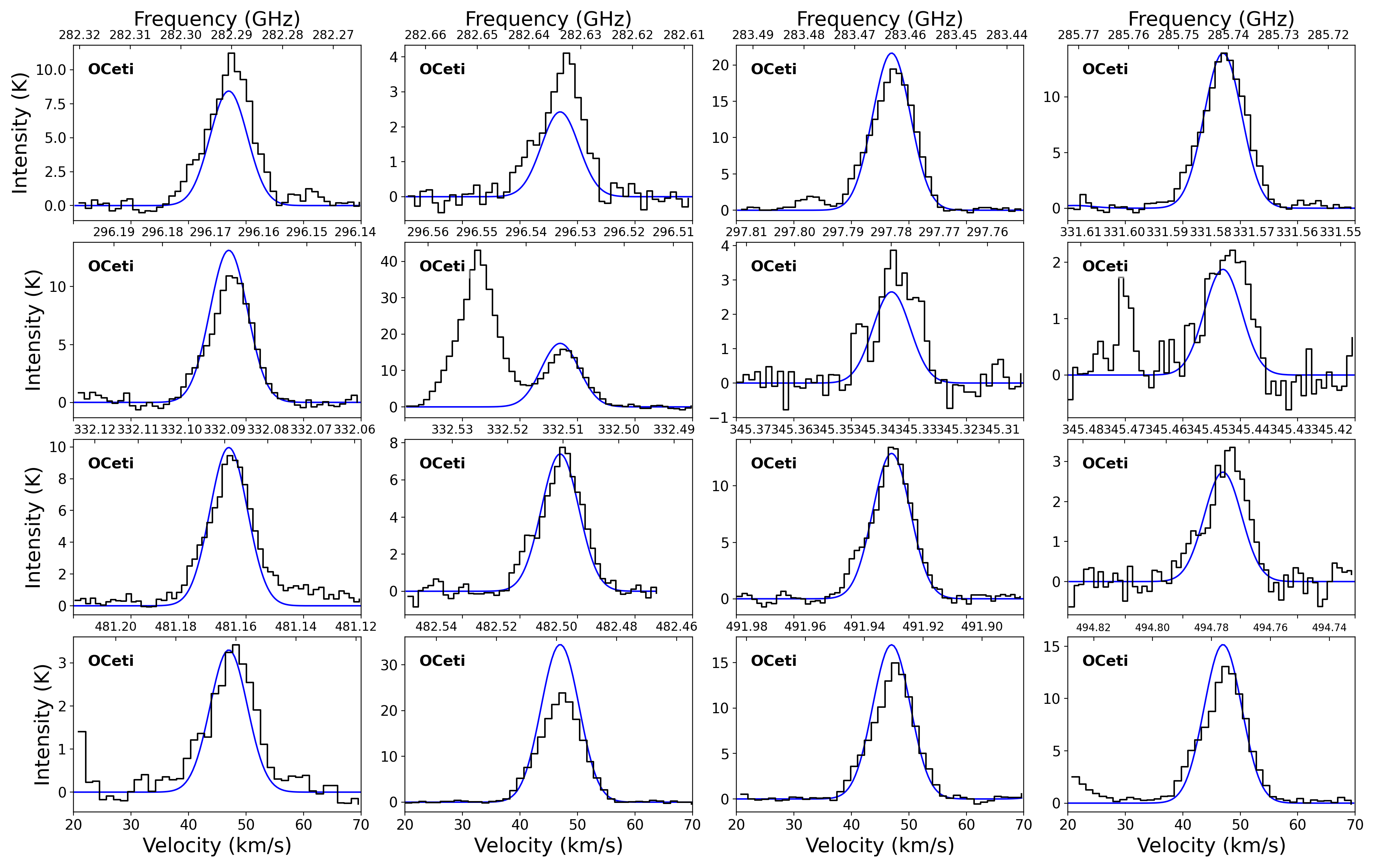}
    \caption{Observed and modeled spectra of SO$_2$ transitions towards $o$ Ceti. The black line represents the observed spectrum, while the blue line represents the modeled spectrum. The strong line at 296.5500 GHz, shown in the second subplot of the third row from the bottom, represents the SO line. The narrow line in the last subplot of the same row corresponds to TiO$_2$ (331.5996 GHz).}
    \label{fig:synth-Oceti}
\end{figure*}

Data calibration and cleaning were performed following standard ALMA procedures using the Common Astronomy Software Application (CASA) \citep{mcmu07}. The uvcontsub task was used to separate the continuum part from the line emissions. Calibration uncertainty depends on the flux calibrator used and typically ranges from 5$\%$ to 20$\%$ \citep{Francis20}. We used CASSIS\footnote{\url{http://cassis.irap.omp.eu}} software for line identification and further analysis \citep{vast15}.

\begin{table*}

\caption{Observation summary}
\label{tab:obs-summary}
\centering{
\begin{tabular}{|cp{1cm}cccccccc|}
\hline
Source &ALMA  &Configuration & No. of &Obs. Date & Ang. res.  & MRS &FOV&  Sensitivity&Spec. Resolution\\
&Band&&Antennas&& (arcsec)&(arcsec)& (arcsec)&(mJy/beam)&kHz  (km s$^{-1}$)\\
\hline 
{\it o} Ceti &8&12m& 46&28-11-18&0.172& 3.212 & 11.96& 0.16&1938.48 ($\sim$1.2)\\
&7&12m& 44&19-11-18&0.212&3.142& 20.08& 0.05&1128.91 ($\sim$1.1)\\
&7&12m& 46&26-11-18&0.213&3.033& 17.21& 0.06&1128.91 ($\sim$1.0)\\
R Leo &8&ACA& 9&24-03-19&2.570&15.118& 20.50& 2.39&1938.48 ($\sim$1.2)\\
&7&12m& 44&06-10-19&0.218&3.288& 16.76& 0.05 &1128.91 ($\sim$1.0)\\
W Hya &8&ACA& 9&08-06-19&2.039&13.634& 20.50& 2.17&1938.48 ($\sim$1.2)\\
&6&12m& 44&08-07-17&0.156&2.117& 22.43& 0.02&976.56 ($\sim$1.1)\\
R Dor &8&ACA& 9&08-06-19&2.135&13.812& 20.50& 1.79&1938.48 ($\sim$1.2)\\
&7&12m& 44&15-09-18&0.194&2.993& 16.05& 0.07&1128.91 ($\sim$1.1)\\
Ep Aqr&8&ACA& 9&24-06-19&2.452&13.939& 20.50& 1.35&1938.48 ($\sim$1.2)\\
\hline
\end{tabular}}
\end{table*}

\begin{figure*}
    \includegraphics[width=0.49\textwidth]{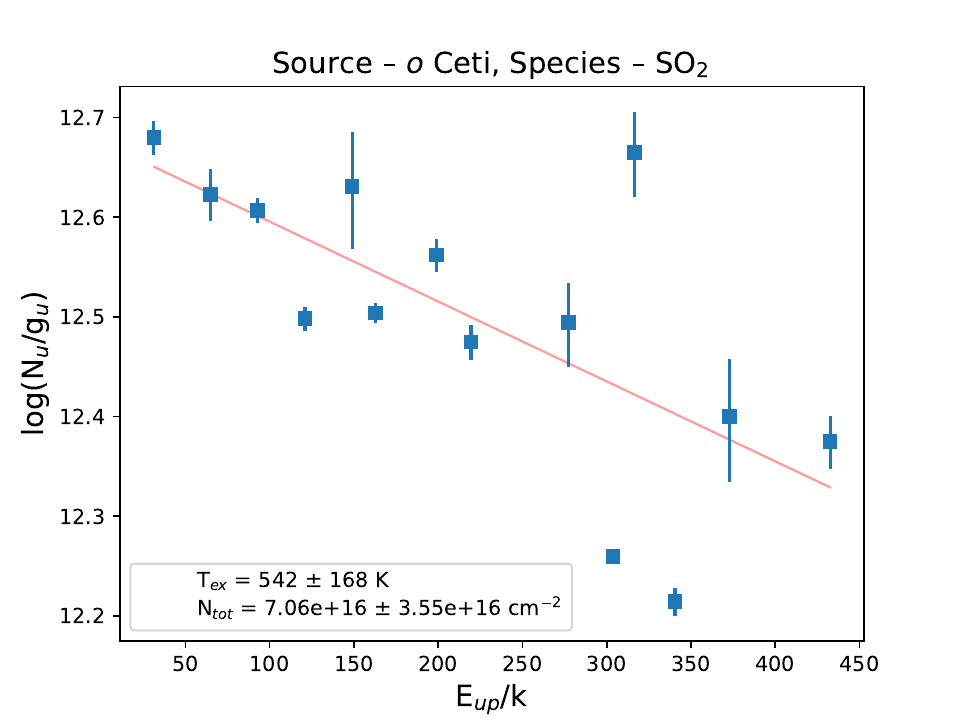}
    \includegraphics[width=0.49\textwidth]{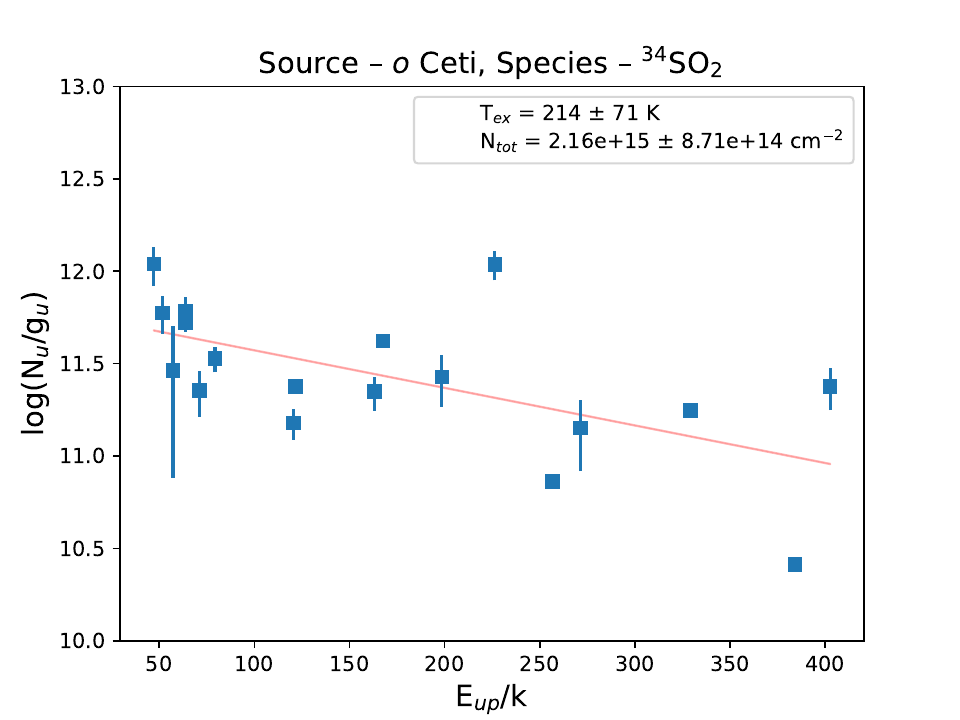}
    \caption{Rotational diagram of SO$_2$ and $^{34}$SO$_2$ for {\it o} Ceti. Blue filled squares represent the data points, and the vertical lines on each data point indicate the error bars. The solid red lines represent the fitted line.}
    \label{fig:rot-diag-oceti}
\end{figure*}

\begin{figure*}
    \includegraphics[width=0.49\textwidth]{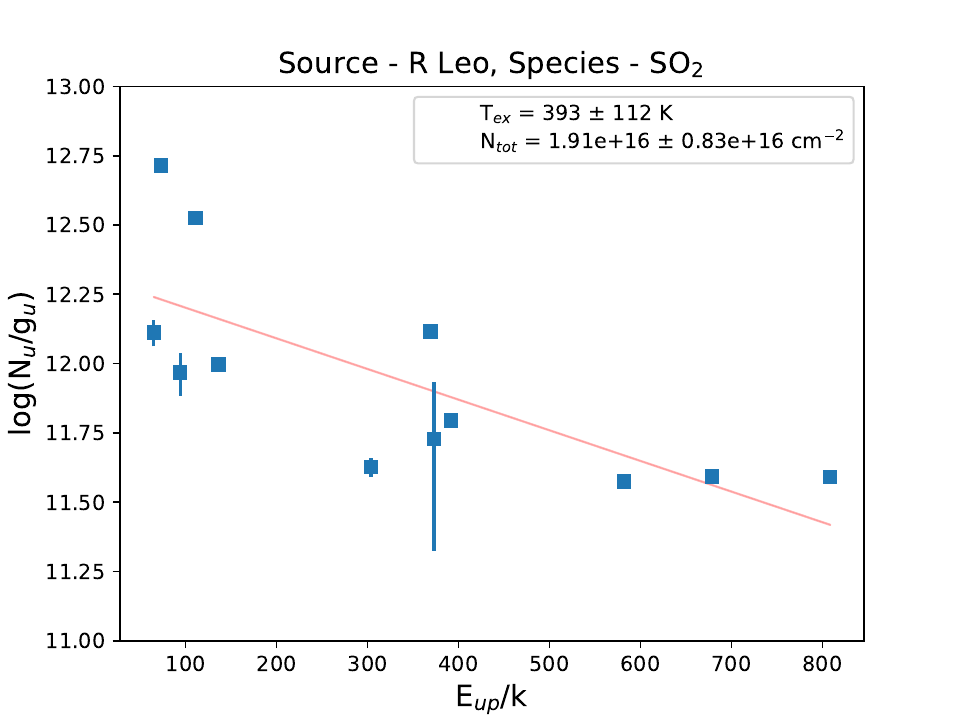}
    \includegraphics[width=0.49\textwidth]{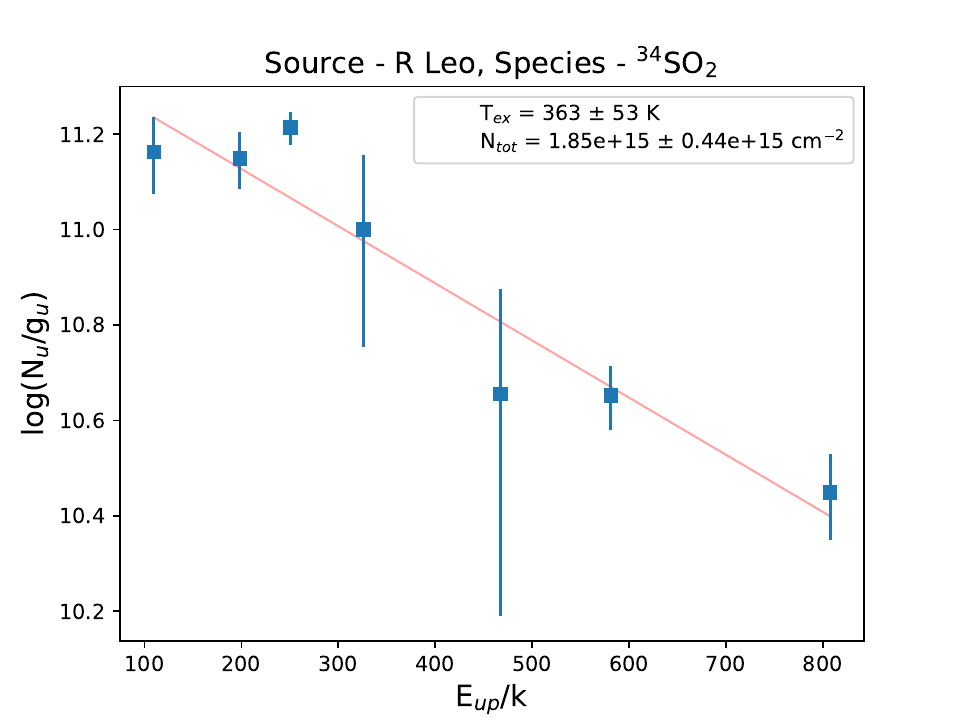}
    \caption{Rotational diagram of SO$_2$ and $^{34}$SO$_2$ for R Leo. Blue filled squares represent the data points, and the vertical lines on each data point indicate the error bars. The solid red lines represent the fitted straight line.}
    \label{fig:rot-diag-rleo}
\end{figure*}

\begin{figure}
    \includegraphics[width=\linewidth]{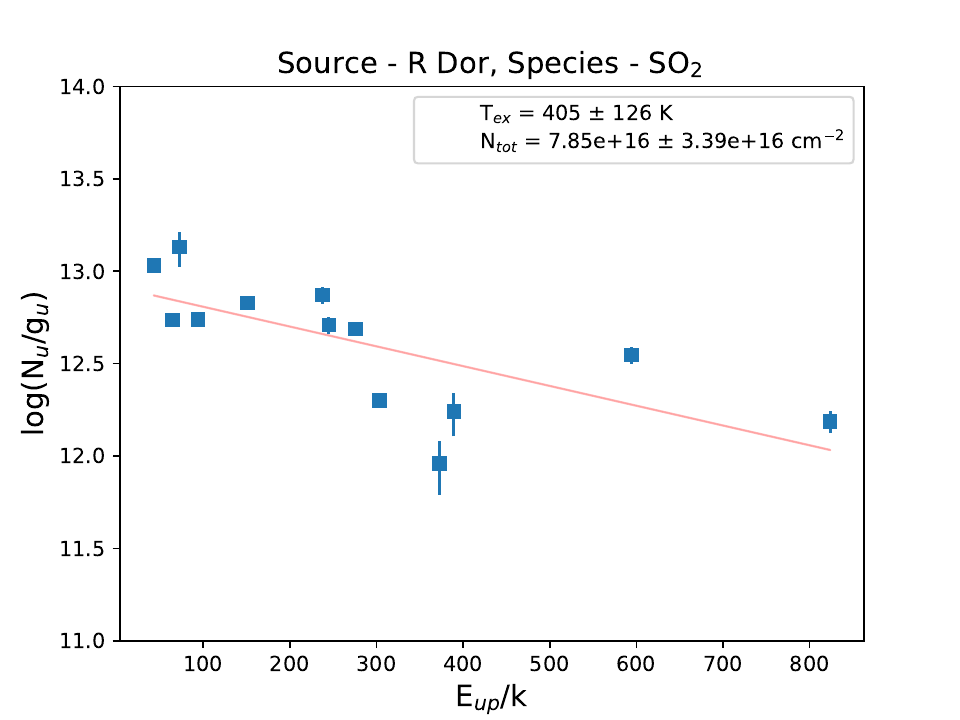}
    \caption{Rotational diagram of SO$_2$ for R Dor. Blue filled squares represent the data points, and the vertical lines on each data point indicate the error bars. The solid red line represents the fitted straight line.}
    \label{fig:rot-diag-rdor}
\end{figure}

\begin{figure}
    \includegraphics[width=\linewidth]{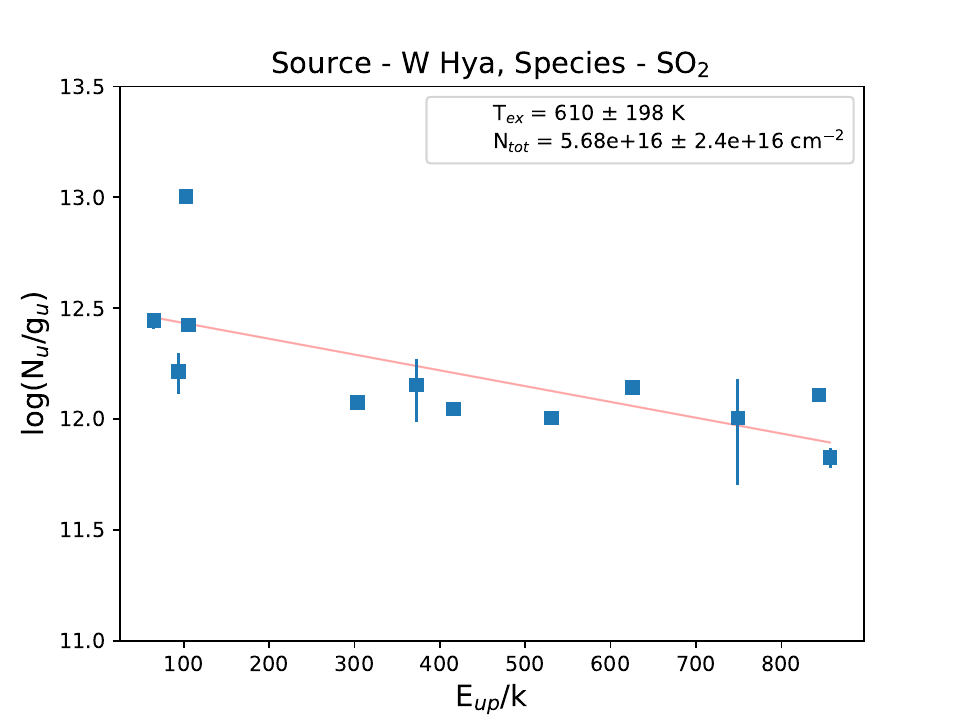}
    \caption{Rotational diagram of SO$_2$ for W Hya. Blue filled squares represent the data points, and the vertical lines on each data point indicate the error bars. The solid red line represents the fitted straight line.}
    \label{fig:rot-diag-whya}
\end{figure}

\begin{figure}
    \includegraphics[width=\linewidth]{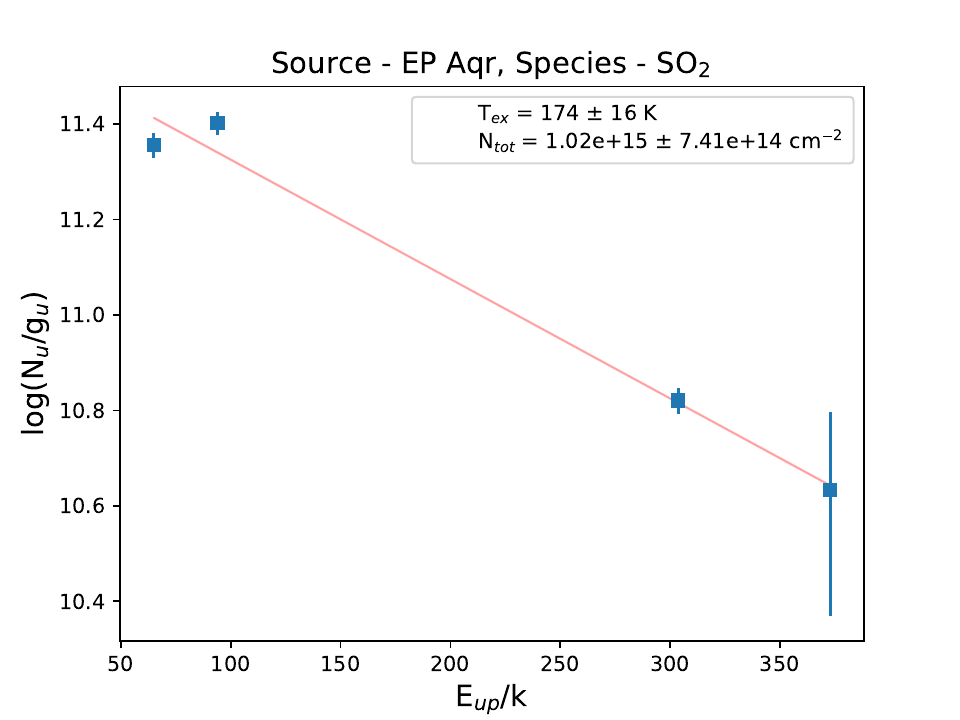}
    \caption{Rotational diagram of SO$_2$ for EP Aqr. Blue filled squares represent the data points, and the vertical lines on each data point indicate the error bars. The solid red line represents the fitted straight line.}
     \label{fig:rot-diag-epaqr}
\end{figure}

\subsection{Line identification}
The line identification of all the observed species was carried out using CASSIS software together with the Cologne Database for Molecular Spectroscopy  \citep[CDMS,][]{mull01,mull05}\footnote{\url{https://www.astro.uni-koeln.de/cdms}} database.
First, we apply a cut based on the rms noise for all data, then we consider the spectral profiles that are above 3$\sigma$, where $\sigma$ is the rms noise. To firmly identify a molecular transition corresponding to the observed spectra, we checked line blending, systematic velocity of the source ($\rm{V_{LSR}}$), and the Einstein coefficient of different nearby transitions.  We choose a few specific transitions (if multiple potential lines are available) corresponding to a particular spectral peak. Finally, to firmly assign a molecular species to the observed spectral feature, we used the Local Thermodynamic Equilibrium (LTE) model using  Markov chain Monte Carlo (MCMC) fitting within CASSIS (see Sec. \ref{sec:mcmc}) and checked whether the synthetic spectrum of a particular species can reproduce the observed profile or not.

\subsection{Markov chain monte carlo (MCMC) fitting \label{sec:mcmc}}
\vskip -0.3cm
 MCMC fitting was employed to fit the observed line profiles of different transitions toward a sample of 5 AGB stars across three different ALMA Bands. For fitting,  we assume that the excitation temperature is equivalent to the gas temperature under LTE conditions. We used the Python scripting interface available in CASSIS for our model calculations to determine the best-fit physical parameters for the SO$_2$ and $\rm{^{34}SO_2}$ transitions. We varied input parameters such as excitation temperature, line width, and column density while keeping the emission regions of the observed transitions, and V$_{LSR}$ fixed (see Table \ref{tab:targets}). After applying the LTE approximation using the MCMC approach, we extracted the best-fit physical parameters such as column density, excitation temperature, and FWHM. We applied the $\chi^{2}$ minimization process to determine the best-fit model that matches the observed line profiles.

\subsection{Rotation diagram analysis \label{sec:rotdia}} 
ALMA observations of our AGB sample contain multiple transitions of SO$_2$ and $\rm{^{34}SO_2}$. Hence, rotation diagram analysis is carried out to determine the excitation temperatures and column densities. Assuming the observed transitions of these species are optically thin and are in LTE, we performed rotational diagram analysis. To check the optically thin assumption, we estimated the optical depth ($\tau$) of each molecular line following the equation:

\begin{equation}
\tau = -ln \Bigg(1-\frac{T_{peak}}{\phi(J(T_{ex})- J(T_{bg})}\Bigg),
\end{equation}
where T$_{peak}$ is the peak emission (peak flux in K from Table \ref{tab:lines}, $\phi$ is the beam filling factor, assumed to be 1, and 

\begin{equation}
{J_\nu(\nu, {T_{ex}}) = \Bigg(\frac{h \nu}{k}\Bigg)\frac{1}{\exp\left(\frac{h \nu}{k {T_{ex}}}\right) - 1}}, 
\end{equation}
where $\nu$  is the frequency of the molecular transition, T$_{ex}$ is obtained from the rotational diagram of each individual source, T$_{bg}$ = 2.73 K. The calculated values of $\tau$ are provided in the last column of Table \ref{tab:lines}. All values of $\tau$ are $<<$ 1, which indicates that the lines are optically thin. For optically thin lines, column density can be expressed as \citep{gold99},
\begin{equation}
\frac{N_u^{thin}}{g_u}=\frac{3k_B\int{T_{mb}dV}}{8\pi^{3}\nu S\mu^{2}},
\end{equation}
where g$_u$ is the degeneracy of the upper state, k$_B$ is the Boltzmann constant, $\rm{\int T_{mb}dV}$ is the integrated intensity, $\nu$ is the rest frequency, $\mu$ is the electric dipole moment, and S is the transition line strength. Under LTE conditions, the total column density can be written as,
\begin{equation}
\frac{N_u^{thin}}{g_u}=\frac{N_{total}}{Q(T_{rot})}\exp(-E_u/k_BT_{rot}),
\end{equation} 
where $T_{rot}$ is the rotational temperature, E$_u$ is the upper state energy, ${Q(T_{rot})}$ is the partition function at rotational temperature. Equation 2 can be rearranged as,

\begin{equation}ln\Bigg(\frac{N_u^{thin}}{g_u}\Bigg)=-\Bigg(\frac{E_u}{k_BT_{rot}}\Bigg)+ln\Bigg(\frac{N_{total}}{Q(T_{rot})}\Bigg).\end{equation}

For the rotational diagram analysis, we combined the observed transitions from ACA and 12m array observations whenever they were available. Since the resolution of the ACA differs from that of the 12m array, we applied a beam dilution factor. We examined the emitting region of both SO$_2$ transitions in ACA and 12m observations and found that the emission is unresolved in our ACA observations. On average, we extracted the spectra with a diameter of 1$^{''}$ for 12m array and 3$^{''}$ for ACA. Hence, we applied a scaling factor of 9 to estimate the upper-level column density of transitions observed with ACA. Finally, we used all data points in the rotational diagram analysis to obtain the rotational temperatures and column densities.  We applied a rotational diagram (for species with more than 2 lines observed) and MCMC fitting to estimate the excitation temperature and column density. The error bars in the rotational diagram come from the Gaussian fitting error of the line profiles. 

To investigate the distribution of SO$_2$ and $\rm{^{34}SO_2}$ in the oxygen-rich AGB stars, we created moment 0 maps of all transitions from both ACA and 12m array observations. Notably, the resolution of our ACA data cannot resolve the emission of SO$_2$, whereas 12m array observations have sufficient resolution to resolve the sulfur species spatially. 

\section{Results \label{sec:3}}
Figure \ref{fig:synth-Oceti} depicts the observed and modeled spectra towards $o$ Ceti. The spectra of other sources are provided in the Appendix (see Figs. \ref{fig:synth-rdor-rleo} and \ref{fig:synth-whya-epaqr}). Table \ref{tab:lines} summarizes all the detected transitions. The line width (Full width at half maximum, FWHM), $\rm{\Delta V}$), $\rm{V_{LSR}}$, and the integrated intensity ($\rm{\int T_{mb} dv}$) of each transition are measured by applying a single Gaussian to fit the observed spectrum. The line parameters of all the observed transitions such as rest frequency ($\nu_0$), quantum numbers ({${\rm J^{'}_{K_a^{'}K_c^{'}}}$-${\rm J^{''}_{K_a^{''}K_c^{''}}}$}), FWHM, $\rm{\int T_{mb} dv}$, upper state energy ($\rm{E_{up}}$), and $\rm{V_{LSR}}$ are noted in Table \ref{tab:lines}.  

The rotational diagrams of SO$_2$ and $\rm{^{34}SO_2}$ (when available) are shown in Figs. \ref{fig:rot-diag-oceti}, \ref{fig:rot-diag-rleo}, \ref{fig:rot-diag-rdor}, \ref{fig:rot-diag-whya}, and \ref{fig:rot-diag-epaqr} for $o$ Ceti, R Leo, R Dor, W Hya, and EP Aqr, respectively.
 Table \ref{tab:temp-colden} compares the derived excitation temperatures and column densities from both RD and MCMC methods.
The results of individual sources are discussed in the following sub-sections. 

\subsection{{\it o} Ceti} 
Many transitions of SO$_2$ (16 lines), $\rm{^{34}SO_2}$ (14 lines), SO (3 lines), and $\rm{^{34}SO}$ (2 lines) are detected with a wide range of upper state energy ($\rm{E_{up}\sim 40 - 700\ K}$; see Table \ref{tab:lines}). Additionally, we detected one transition of $\rm{^{33}SO}$ (343.0861 GHz). We also find a line close to the frequency of a $\rm{^{33}SO_2}$ line (345.5852 GHz), but do not associate it with $\rm{^{33}SO_2}$ emission because other lines potentially expected to be seen (at 283.338 GHz, 296.270 GHz, 334.030 GHz, and 479.960 GHz) are not observed. We have multiple transitions of SO$_2$ and $\rm{^{34}SO_2}$ , hence applied rotation diagram analysis (see Figs. \ref{fig:rot-diag-oceti}). The rotational diagram analysis yields excitation temperatures and column densities of ${T_{\rm ex}}= 542\pm 168$ K and ${N_{\rm SO_2} = (7.06\pm3.55)\times10^{16}}$ $\rm{cm^{-2}}$ for SO$_2$ and ${T_{\rm ex}}= 214\pm 71$ K and ${N_{\rm ^{34}SO_2} = (2.16\pm0.87)\times10^{15}}$ $\rm{cm^{-2}}$ for $\rm{^{34}SO_2}$. Our estimated temperature of SO$_2$ is comparable with the previous measurement based on TiO (474 $\pm$ 69) K \citep{Kaminski2017} and AlF observations within uncertainties \citep{saberi2022}. For {\it o} Ceti, a very faint signal at 493.3622 ($E_{up}\approx564$~K) GHz transition of H$_2$S is observed toward the central source, but it is below the detection limit. Assuming LTE, we created synthetic spectra to match the observed profile. For this, we adopted the same temperature of 542~K, as obtained from the RD of {\it o} Ceti, and estimated an upper limit for the column density of H$_2$S to be $6.53\times10^{14}$\ cm$^{-2}$. However, we did not detect any signal of the high-excitation transition ($E_{up}\approx564$~K) toward the arc-like structure of {\it o} Ceti. \cite{Danilovich2017} found that H$_2$S likely plays an important role in oxygen-rich AGB stars with high mass-loss rates, but is unlikely to be significant in stars of other chemical types or in oxygen-rich stars with lower mass-loss rates.

Figures \ref{fig:mom0-octei-band7}-\ref{fig:mom0-oceti-band8} show the moment 0 maps of all detected transitions (also see Fig. 1 in the supplementary document). Spatial analysis with high-resolution ($\sim 0.2^{''}$) observations with different ALMA bands show centralized emission for all transitions with emitting regions around $\sim$ 1$^{''}$ towards the main continuum source. The shape of the emission shows irregular appearances,  which could be due to the binary companion. The average separation between binary sources is 0.472$^{''}$ \citep{Vlemmings2015}, corresponding to $\sim$48 au, considering a distance of 102 pc. Nonetheless, we cannot distinguish the distinct features of SO and SO$_2$ around Mira A and B with the present angular resolution. 

Furthermore, only for several transitions ($\rm{ E_{up} < 200\ K}$), we find extended emission in the south-west position, which is completely offset and is not connected with the emission that is close to the continuum peak. For example, the low-excitation transitions of SO and SO$_2$ (see Figs. \ref{fig:channel-maps-oceti-SO} and \ref{fig:channel-maps-oceti-SO2}) show an extended emission, about 3$^{''}$ ($\sim$ 300 au) offset from the continuum position. Because of the lower intensity of the offset emission, it does not appear as a prominent feature in the integrated intensity maps. Hence, we plotted channel maps for these transitions, where the extended emission is more clearly visible. In these figures, we see the extended feature in a few channels around $\pm$3 km s$^{-1}$ of the V$_{LSR}$ = 47 km s$^{-1}$. Previously, \cite{Wong2016} reported the extended spatially resolved emission in SiO ($J=$5-4) ALMA observation around the same position. Our observations show SO and SO$_2$ emissions around the same region where SiO appears. Since SiO is a good tracer of shocks \citep{Fontani2019}, the formation of gas-phase SO and SO$_2$ is probably caused by shock-induced processes in the same environment.
It could also be a region with higher gas density. The morphological correlation between SO, SO$_2$, and SiO emissions implies that the shock processes might enhance the gas-phase abundance of these molecules. Shocks can significantly influence chemistry by increasing temperatures and densities and disrupting dust grains. In particular, shock waves can sputter dust grains and ice mantles \citep{Suutarinen2014}, releasing sulfur-bearing species such as atomic sulfur (S), hydrogen sulfide (H$_2$S), or sulfur monoxide (SO) into the gas phase. These species can then undergo rapid gas-phase reactions to form  SO$_2$. Furthermore, we estimated that the temperature of this region is $\sim133 \pm 20$ K, based on an LTE analysis of all detected transitions of SO and SO$_2$ towards the offset position (see Table \ref{tab:lines-oceti-arc}.

\begin{table*}
\caption{Rotational temperature and column density of SO$_2$ and $\rm{^{34}SO_2}$ from two methods}
    \centering
    \begin{tabular}{|c|c|c|c|c|c|c|c|c|c|}
    \hline
    Source & \multicolumn{2}{|c|}{$T_{\rm ex} ({\rm{{SO_2}}}$) (K)} &\multicolumn{2}{|c|} {$\rm{N ({SO_2})}$}  &\multicolumn{2}{|c|}{$T_{\rm ex}$($\rm{{^{34}SO_2}}$) (K)} &\multicolumn{2}{|c|} {$\rm{N ({^{34}SO_2})}$}&{$\rm{N_{SO_2}}$/$\rm{N_{^{34}SO_2}}$} \\
    \hline
    & {RD} & {MCMC} & {RD} & {MCMC} & {RD} & {MCMC} & {RD} & {MCMC} & \\
    \hline
    {{\it o} Ceti} & 542$\pm$168 & 410$\pm$36 & 7.06e+16 & 6.76e+16 & 214$\pm$71 &296$\pm$4& 2.16e+15 & 2.21e+15 & 32$\pm$6\\
    { R Leo} & 393$\pm$112 &489$\pm$29 & 1.91e+16 &2.49e+16& 363$\pm$53 &436$\pm$48 & 1.85e+15 &1.89e+15& 10$\pm$3\\
    { R Dor} & 405$\pm$126 & 364$\pm$26 & 7.85e+16 &4.05e+16 & -- & 258$\pm$54 &--  &1.55e+15& 26$\pm$8  \\
    { W Hya} & 610$\pm$198 & 419$\pm$4 & 5.68e+16 &5.54e+16& -- & 573$\pm$22 &--  &3.17e+15 & 18$\pm$4 \\
    { EP Aqr} & 174$\pm$16 & 209$\pm$4 & 1.02e+15 & 1.97e+15 & -- & 176$\pm$37 & -- & 1.61e+14 & 12$\pm$5\\
    \hline
    \end{tabular}
    \label{tab:temp-colden}\\
Note: RD- Rotational Diagram Method, MCMC- Monte Carlo Markov Chain Method. For MCMC fitting, we used the following emission regions: {\it o} Ceti- 1$^{''}(\sim $102 au), R Dor- 1.5$^{''}(\sim $68 au), R Leo 1$^{''}(\sim $130 au), W Hya - 1$^{''}(\sim $104 au), and EP Aqr - 3$^{''}(\sim $330 au), which are obtained from the moment 0 maps analysis.
\end{table*}

\subsection{R Dor}
A list of detected transitions of SO$_2$ (14 lines), $\rm{^{34}SO_2}$ (2 lines), SO (1 line), and $\rm{^{34}SO}$ (1 line) towards R Dor is provided in Table \ref{tab:lines}. We applied rotational diagram analysis for SO$_2$ to obtain the excitation temperature (see Fig. \ref{fig:rot-diag-rdor}). Our derived excitation temperature ${T_{\rm ex}}$= 405$\pm$126 K and ${N_{\rm SO_2} = (7.85\pm 3.39)\times10^{16}}$ $\rm cm^{-2}$. We do not have multiple transitions of $\rm{^{34}SO_2}$  and SO. Hence, we measured the excitation temperature and the column density of these species using the MCMC method. Our estimated excitation temperature using SO$_2$ is consistent with previous measurements $\sim 505\ K$ \citep{khouri2024}. Moment 0 maps of different transitions of SO$_2$ are in Fig. \ref{fig:mom0-Rdor-12m} (also see Fig. 2 in the supplementary document). It is clear from Fig. 2 in the supplementary document that SO$_2$ is unresolved with ACA observation, but with high-resolution data, it is well resolved and shows clumpy emission structures (see Fig. \ref{fig:mom0-Rdor-12m}) where the continuum peak is at the center.  Furthermore, emission shows more compact features for transitions with high upper-state energy than the lower E$_{up}$ transition. Our results, obtained at an angular resolution of $0.2^{''}$, are consistent with previous ALMA observations at a comparable resolution ($\sim 0.15^{''}$) \citep{Danilovich2020}. The co-spatial emission of SO and SO$_2$ (see Fig. \ref{fig:mom0-Rdor-12m}) suggests that both molecules trace the same wind structure within the circumstellar environment. Furthermore, since the MRS of our observations is approximately 3$^{''}$, we compared the observed intensities of transitions with similar $E_{\rm up}$ between our data and previous results obtained using the APEX 12m single-dish telescope \citep{Danilovich2016}. This comparison was performed to determine whether any flux has been resolved out in the interferometric data. However, we found good agreement between the single-dish and interferometric results for transitions with comparable $E_{\rm up}$.

\subsection{R Leo}
All detected transitions of SO$_2$ (10 lines), $\rm{^{34}SO_2}$ (7 lines), SO (1 line), and $\rm{^{34}SO}$ (1 line) towards R Leo are summarized in Table \ref{tab:lines}. We have multiple transitions of SO$_2$ and $\rm{^{34}SO_2}$. Hence, we applied a rotational diagram to estimate the excitation temperature and column density (see Figs. \ref{fig:rot-diag-rleo}). The rotational diagram analysis yields excitation temperatures and column densities of $\rm{T_{ex}}= 393\pm112$ K and ${N_{\rm SO_2} = (1.91\pm0.83)\times10^{16}}$ $\rm{cm^{-2}}$ for SO$_2$ and ${T_{\rm ex}}= 363\pm53$ K and ${N_{\rm ^{34}SO_2} = (1.85\pm0.44)\times10^{15}}$ $\rm{cm^{-2}}$ for  $\rm{^{34}SO_2}$. For SO, we do not have multiple transitions. Consequently, we employed the MCMC method to derive a column density of SO of $\rm{6.3\times10^{15}\ cm^{-2}}$ and an excitation temperature of $\rm{T_{ex}}$ = 300 K. Figures \ref{fig:mom0_RLeo_12m}, and \ref{fig:mom0_RLeo_so-34so-34so2} (also see Fig. 3 in the supplementary document) show moment 0 maps of all detected transitions of SO$_2$, $\rm{^{34}SO_2}$, and SO and $\rm{^{34}SO}$. It is clear from Fig. 3 in the supplementary document that emission of SO$_2$ is not resolved with a low-resolution ($\sim 2.6^{''}$) observation. In contrast, SO$_2$ can be resolved well with high-resolution ($\sim 0.22^{''}$) data as shown in Figs. \ref{fig:mom0_RLeo_12m} and \ref{fig:mom0_RLeo_so-34so-34so2}. We find a morphological correlation in the emission between SO$_2$, $\rm{^{34}SO_2}$, and SO. The emission from the low-excitation transitions of ($E_{\rm up}$ = 81 K) and $\rm{^{34}SO_2}$ ($E_{\rm up}$ = 77 K) appears slightly more extended compared to higher-excitation transitions. Overall, the emission from both SO and SO$_2$ is centrally concentrated and co-spatial, suggesting that these species likely trace the same wind structures in the CSE of R Leo.

\subsection{W Hya}
The observed transitions of SO$_2$ (13 lines) and $\rm{^{34}SO_2}$ (3 lines) towards W Hya are noted in Table \ref{tab:lines}.  The estimated excitation temperature using SO$_2$ rotational diagram is ${T_{\rm ex}}= 610\pm198$ K and column density ${N_{\rm SO_2} = (5.68\pm2.4)\times10^{16}}$ $\rm{cm^{-2}}$.  Moment 0 maps for different transitions of SO$_2$ with high- ($\sim 0.16^{''}$) and low-resolution  ($\sim 2.0^{''}$) observations are depicted in Fig. \ref{fig:mom0_Whya_12m} and Fig. 4 in the supplementary document, respectively. In the case of W Hya, emission from SO$_2$ is not resolved in the ACA observations, like R Dor and R Leo. However, we find an additional structure of strong SO$_2$ emission located approximately 10$^{''}$ offset from the central continuum peak. It possibly traces the structures at the outer envelope of W Hya. At high resolution, the SO$_2$ emission is resolved. It exhibits an ordered, centrally concentrated structure, with a slight offset between the continuum and molecular emission peak, similar to what is observed in R Leo. However, we do not detect the additional features at the offset position, which we found from ACA observations. The structure at the offset position is most likely resolved out due to the limited MRS of 2.1$^{''}$ in the high-resolution observations. We do not have SO transition in the frequency range of our observation. Nonetheless, modeling results from previous studies suggest that the SO$_2$ distribution is similar in size and abundance to the circumstellar SO distribution \citep{Danilovich2016}.

\subsection{EP Aqr} 
For EP Aqr, we have several transitions of SO$_2$ (5 lines) and one transition of $\rm{^{34}SO_2}$ (1 line) but only with ALMA Band 8 ACA observations (see Table \ref{tab:lines}). The rotational diagram yields ${T_{\rm ex}}$= 174$\pm$16 K and ${N_{\rm SO_2} = (1.02\pm0.74)\times10^{15}}$ $\rm{cm^{-2}}$.  Figure 5 in the supplementary document shows the moment 0 maps of SO$_2$ transitions. The slightly lower temperature obtained compared to other sources (see Table \ref{tab:temp-colden}) could be due to the large emission regions and the lack of high-resolution data with multiple transitions for this source. All transitions are spatially unresolved because of low angular resolution $\sim 2.5''$. \cite{Tuan2019} reported one transition of SO$_2$ and found that the emission is confined toward the center of the source, with an emitting diameter of $\sim 0.25{''}$ ($\sim$30 au). They suggested that SO$_2$ could be a useful tracer of the mass-loss mechanism in its early phase. Their analysis also provided clear evidence of rotation, possibly combined with moderate expansion. Furthermore, by applying the temperature radial dependence from \cite{Hoai2019} and considering the radial extent of the SO$_2$ emission, they estimated a gas temperature of a few hundred kelvin, consistent with values typical for oxygen-rich AGB stars \citep{Yamamura1999}. However, multi-transition (with a wide range of $E_{\rm up}$) with high-angular resolution observations could provide better constraints on both spatial distribution and actual gas temperature traced by SO$_2$.

\begin{figure*}
    \centering
    \includegraphics[width=\textwidth]{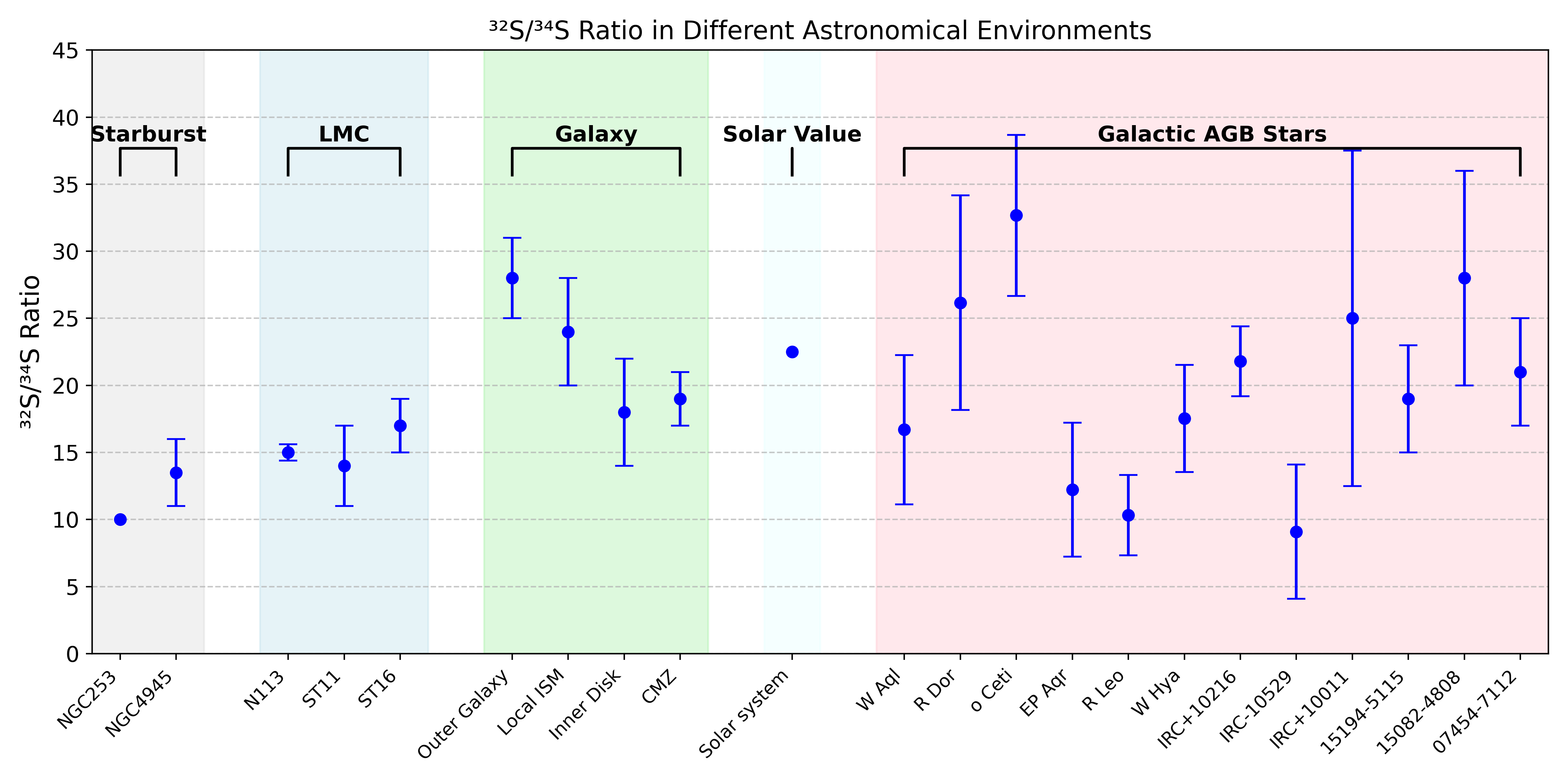}
    \caption{$\rm{^{32}S}$/$\rm{^{34}S}$ ratios for different environments. N113- \cite{Gong2023}; ST11, ST16 - \cite{Shimonishi2020}; NGC 253 - \cite{Martin2021}; NGC 4945 - \cite{Wang2004}; Solar System - \cite{Anders1989}; CMZ, Inner Disk, Local ISM, Outer Galaxy - \cite{Yan2023}; IRC+10216 - \cite{Mauersberger2004}; W Aql, IRC+10011, IRC-10529 - \cite{Wallstrom2024}; 15194-5115, 15082-4808, 07454-7112- \citep{Unnikrishnan2024}; {\it o} Ceti, R Dor, W Hya, R Leo, EP Aqr - This work. Additionally, \cite{Danilovich2020} reported a ratio of 42 for the Galactic AGB star, IK Tau, which is not included in the plot due to the absence of measured uncertainties.}
    \label{fig:32S-34S-ratio}
\end{figure*}

\section{Discussion \label{sec:4}}
\subsection{SO and SO$_2$ distributions}
SO and SO$_2$ have been previously found to have centralized distributions for low mass-loss rate AGB stars and shell-like distributions for high mass-loss sources \cite[e.g.,][]{Danilovich2016,Wallstrom2024}. They also observed that sources with shell-like emission are typically detected in lower-energy SO$_2$ lines, while those with centralized emission are more often detected in higher-energy lines. All five stars in our sample have relatively low mass-loss rates. We investigated the distribution of SO$_2$ and SO toward these sources. For {\it o} Ceti, we observe slightly irregular emission contours for SO$_2$, $\rm{^{34}SO_2}$, SO, and $\rm{^{34}SO}$, and emission peaks are offset from the continuum peak. The emissions of SO$_2$ and $\rm{^{34}SO_2}$ are spatially correlated, as are the emissions of SO and $\rm{^{34}SO}$. However, with the current angular resolution, we can not distinguish the distributions of our detected species between the Mira A and B components. Furthermore, we detected extended emission for several transitions of SO and SO$_2$ and some other species (see Table  \ref{tab:lines-oceti-arc}) at the south-west position, which is $\sim 3''$ offset from the main continuum and molecular emission peaks (e.g., see Figs. \ref{fig:channel-maps-oceti-SO} and \ref{fig:channel-maps-oceti-SO2}). The extended emission region is also seen with the SiO (5-4) transition \citep{Wong2016}, suggesting that shocks may play a significant role in the origin of SO and SO$_2$. In addition, the measured temperature of this arc-like structure is lower than that of the central region. 
Distributions of $\rm{^{34}SO}$ and $\rm{SO_2}$ show a morphological correlation in R Dor. The emissions of these species are clumpy, but the emission peak appears at the center of the continuum peak. For R Leo, we find ordered emission structures of $\rm{SO_2}$ and $\rm{^{34}SO_2}$ centered at the continuum peak. We find similar features for SO and $\rm{^{34}SO}$ but slightly extended compared to $\rm{SO_2}$ and $\rm{^{34}SO_2}$ emission. SO and $\rm{^{34}SO}$ probably trace the cold gas of the outer layer of the CSE. In W Hya, we find SO$_2$ and $\rm{^{34}SO_2}$ contours with an approximate circular pattern without smooth arcs, and emission peaks are slightly offset from the continuum peak. Apart from the central emission around the continuum peak, additional strong emission peaks appear at north-west, which is 10$^{''}$ offset from the continuum peak. SO$_2$ and $\rm{^{34}SO_2}$ are not resolved with the ACA observations of EP Aqr. In summary, our results support the previous conclusion that SO and SO$_2$ exhibit a centralized distribution in low mass-loss rate AGB stars.

Recently \cite{danilovich2025} observed low excitation ($E_{low} < 100\ $K) transitions of SO$_2$ towards intermediate-mass AGB star OH 30.1-0.7. The source is thought to have a very high mass loss rate ($> 10^{-4} M_{\odot}\ yr^{-1}$). They found shell-like distributions of SO$_2$, consistent with morphology as discussed above for sources with high-mass loss rates. Their results suggest a low excitation temperature for SO$_2$ around the source. They also concluded that their result aligns with the observed trend that AGB stars with high-mass loss rates tend to show lower-energy SO$_2$ transitions \citep{Wallstrom2024}. In contrast, we have detected both low and high-excitation transitions in a sample of AGB stars with low mass-loss rates. We obtained high excitation temperatures of SO$_2$ for all sources (see Table \ref{tab:temp-colden}), which are of the order of a few hundred kelvin, indicating that SO$_2$ originates mainly from the intermediate region of the CSE in all sources.  The average high angular resolution of our observations is $\sim$ 0.2$^{\prime\prime}$ and the MRS $\sim$3$^{\prime\prime}$. Since the emitting regions of the detected species are confined to within $\sim$ 2.0$^{\prime\prime}$ (see Figs. \ref{fig:mom0-octei-band7} \ref{fig:mom0-octei-so-34so}, \ref{fig:mom0-oceti-band8}, \ref{fig:mom0-Rdor-12m}, \ref{fig:mom0_RLeo_12m}, \ref{fig:mom0_RLeo_so-34so-34so2}, and \ref{fig:mom0_Whya_12m}), and the MRS is consistently higher than this, we can assume that we recover all the flux. Therefore, we do not expect significant flux loss due to spatial filtering.

The centralized emission of SO and SO$_2$ towards {\it o} Ceti, W Hya, and R Leo could be due to several physical properties of their surrounding environments. \cite{Khouri18} found dust surrounding both Mira A and B, as well as a dust tail connecting the two stars. They found a region around Mira A that exhibits high gas density close to the star, which is characterized by a steep decline in density at its outer edge. Mira is a binary system, and within this context, the observed SO and SO$_2$ emission is centrally concentrated but displays an irregular morphology, suggesting that their formation in the inner envelope might be influenced by both local density enhancements and dynamical effects related to binarity. R Leo has episodic and patchy mass ejection circumstellar environment \citep{Hoai2023} leading to localized shocks in the inner envelope, which in turn may enhance SO and SO$_2$ formation, producing the observed centralized emission. The circumstellar envelope of the stable component of W Hya shows an approximately spherical morphology, with the gas and dust expanding radially \citep{Hoai2022}. The SO and SO$_2$ emissions are centrally concentrated, indicating an origin in the warm, dense inner envelope, where possibly shocks and dust-formation processes predominate in sulfur chemistry.

Ionization effects are likely to play only a minor role in these sources, as the species reside in the inner envelope, where they are expected to be shielded from the interstellar radiation field. Furthermore, the centralized SO and SO$_2$ emissions indicate that these molecules are not primarily produced by outer-shell photochemistry. However, the present data cannot resolve the molecular emission features between the individual components in sources with binary companions. Furthermore, the modeling results of \cite{Danilovich2016} found that, in their sample of low-mass-loss-rate stars, both SO and SO$_2$ exhibit centrally peaked abundance distributions. In contrast, in higher mass-loss rate stars, the SO abundance peaks further out in the wind. The SO$_2$ abundance profile might follow a similar profile \citep{Danilovich2020}. However, dedicated studies of individual sources, combining observations and modeling, are needed to better understand the origin of emission features of sulfur-bearing molecules.

\subsection{Sulfur isotopic ratio in various astrophysical environments}
All sources in our sample are oxygen-rich AGB stars with similar metallicity and mass-loss rates of approximately 10$^{-7}$ $\rm{M_{\odot}\ yr^{-1}}$ (see Table \ref{tab:targets}). We estimated the $\rm ^{32}S/^{34}S$ ratio for $\it{o}$ Ceti and R Leo using column density values that are obtained using RD analysis, and for W Hya, R Dor, and EP Aqr, values obtained using MCMC analysis. Our measured isotopic ratios with their uncertainties are listed in Table \ref{tab:temp-colden}. We estimated $\rm ^{32}S/^{34}S$ ratio of 32$\pm$6 for $o$ Ceti, 10$\pm$3 R Leo, 26$\pm$8 for R Dor, 18$\pm$4 for W Hya, and 12$\pm$5 for EP Aqr. Comparing these ratios with the Solar value of 21.7, our estimated values for R Dor and W Hya are in good agreement with the solar value. The ratios for EP Aqr and R Leo are slightly lower, and for {\it o} Ceti, it is slightly higher than the expected value. We also measured $\rm ^{33}S/^{34}S$ ratio $\sim$ 0.35 for {\it o} Ceti, which is higher than solar value \citep[$\sim$ 0.17,][]{Asplund21} by a factor 2. However, this is only in one source and based on one line of $\rm{^{33}SO}$. Hence, we need multitransition observations to constrain this ratio and to compare this with different astrophysical environments.

Along with SO$_2$ and $\rm{^{34}SO_2}$ transitions, we also have one pair of SO and $\rm{^{34}SO}$ detection towards $o$ Ceti, R Leo, and R Dor. In the case of $o$ Ceti, an approximate $\rm{^{32}SO}$/$\rm{^{34}SO}$ ratio of 7$\pm$1 is obtained using the integrated intensities of transitions with similar $E_{\rm up}$ and transition probabilities. This is considerably lower than the ratio of 32.4 obtained from SO$_2$ isotopologue ratio. This can be partly due to opacity effects, which we did not take into account in our calculations. The emitting regions of two molecules may also differ, leading to over- or underestimation of the column densities and derived isotopic ratios. For R Dor, we have one transition of SO with high E$_{up}$ = 143 K and $\rm{^{34}SO}$ with low E$_{up}$ = 80 K. Also, their transition are very different. Hence, we did not estimate the $\rm{^{32}SO}$/$\rm{^{34}SO}$. Our measured value of $\rm{^{32}SO}$/$\rm{^{34}SO}$ is 6.0$\pm$0.5 for R Leo, which is obtained using integrated intensities of two transitions with similar E$_{up}$ and transition probability. This value is also considerably lower than the value of 26 obtained from the SO$_2$ isotopologue ratio, likely for the same reason as discussed for $o$ Ceti.

Our measured value of $\rm{^{32}SO_2}$/$\rm{^{34}SO_2}$ towards two oxygen-rich AGB stars R Dor and W Hya agrees well with the solar value within uncertainties, indicating that $\rm{^{32}S}$ and $\rm{^{34}S}$ are not significantly processed during the AGB phase. However, the measured ratio for $o$ Ceti is slightly higher and for R Leo and EP Aqr are lower. The obtained value of $\rm{^{32}S/^{34}S}$ ratio in R Leo and IRC-10529 is significantly lower compared to solar value, which can be attributed to the fact that these sources might have lower metallicity or formed in an environment with lower $\rm{^{32}S/^{34}S}$. Another possibility could be the age effect of the different environments. Conversely, our estimated $\rm{^{32}SO}$/$\rm{^{34}SO}$ ratio deviates from the solar value, which could be attributed to the SO line being optically thick. In such cases, the observed line intensity underestimates the true SO column density, leading to an artificially low isotopic ratio. To verify the accuracy of the $\rm ^{32}S/^{34}S$ ratio obtained here, it would be useful to derive the sulfur isotopic ratio using other sulfur-bearing species abundant in the CSEs of oxygen-rich AGB stars, such as CS and SiS, in addition to SO and SO$_2$, as previously performed by \cite{VelillaPrieto17} for IK Tau.

Furthermore, we compare our derived $\rm{^{32}S/^{34}S}$ ratios with different astrophysical environments (see Fig. \ref{fig:32S-34S-ratio}), such as the Large Magellanic Cloud (LMC), representing a low-metallicity environment with a metallicity of $\sim 0.3-0.5\ Z_\odot$ \citep{Westerlund1997}; starburst galaxies, NGC 253 with metallicity $\sim 0.5-1.5\ Z_\odot$ \citep{Beck2022}; NGC 4945 with metallicity $\sim 0.5-1.2\ Z_\odot$ \citep{Mouhcine2005,Stanghellini2015}; Milky Way (central part has solar metallicity and decreasing metallicity with distance), considering different environments; outer Galaxy has metallicity similar to LMC \citep{Shimonishi2021}, the local ISM; the inner disk; the CMZ; the solar system; and finally, a list of AGB stars that includes our work and some literature values. It is clear from Figure \ref{fig:32S-34S-ratio} that the $\rm{^{32}S/^{34}S}$ ratios for most of the Galactic AGB stars except R Leo and IRC-10529, are comparable to the solar value, including different environments from our Galaxy and Solar system. Since $\rm{^{32}S}$ and $\rm{^{34}S}$ are primarily produced through explosive nucleosynthesis during Type II and Type Ia supernovae, the measured values in AGB stars reflect the abundances in their natal clouds. In contrast, $\rm{^{32}S/^{34}S}$ ratios in LMC and starburst galaxies are lower compared to the Solar value. Based on comprehensive calculations, \cite{Woosley95} reported that $\rm{^{32}S}$ and $\rm{^{33}S}$ are the primary yield (in the sense that the stellar yields do not strongly depend on the initial metallicity of the stellar model) and do not depend on the initial metallicity of the stellar yield i.e., metal content of a star at its birth, while $\rm{^{34}S}$ is not a clean primary isotope, and its yield decreases with metallicity. Recently, \cite{Gong2023} also reported $\rm{^{32}S}$/$\rm{^{33}S}$ and $\rm{^{34}S}$/$\rm{^{33}S}$ in the LMC are lower than the Milky Way and NGC 253 and concluded which can be attributed to combination of age (may be composed of older stellar populations that have not undergone as much recent nucleosynthesis), low metallicity (influences the types and outcomes of nuclear reactions in stars), and star formation history (e.g., rate of star formation). 

The $\rm{^{32}S/^{34}S}$ isotopic ratio shows an increasing trend from NGC 253 (youngest) to the Milky Way (oldest) (see Fig.~\ref{fig:32S-34S-ratio}), which may correlate with the age of stellar populations across these galaxies. NGC 253, a starburst galaxy with the lowest $\rm{^{32}S/^{34}S}$ ratio, is dominated by young, massive stars due to intense recent star formation, which may indicate minimal isotopic processing. NGC 4945, with a slightly higher $\rm{^{32}S/^{34}S}$ ratio, has a mix of younger and older stars, reflecting a moderately older stellar population than NGC 253. The LMC with an even higher $\rm{^{32}S/^{34}}S$ ratio, contains stars several billion years old, making it older than NGC 4945 but younger than the Milky Way. The Milky Way, with the highest $\rm{^{32}S/^{34}S}$ ratio, hosts the oldest stellar populations, including stars over 13 billion years old, suggesting significant nucleosynthetic evolution. This trend indicates that lower $\rm{^{32}S}$/$\rm{^{34}S}$ ratios are associated with younger, less processed stellar populations. In contrast, higher ratios correspond to older, more evolved galactic environments where sulfur chemistry has been processed significantly. However, this analysis has been based on a small sample. Therefore, we need a large sample and systematic analysis to confirm this trend with multi-transition different sulfur-bearing species along with different isotopic ratios such as  $\rm{^{33}S/^{34}S}$, $\rm{^{32}S/^{33}S}$, and $\rm{^{32}S}$/$\rm{^{34}S}$.

\section{Conclusions \label{sec:5}}

We have analyzed the data of five oxygen-rich AGB stars with low-resolution ($\sim 2.5''$) and complementary high-resolution ($\sim 0.2''$), obtained with ALMA ACA and 12m array. The major conclusions of this paper are summarized as follows.

\begin{itemize}
\item We report the detection of 16 transitions of $\rm{SO_2}$, 14 of $\rm{^{34}SO_2}$, 3 of $\rm{SO}$, and 2 of $\rm{^{34}SO}$ toward {\it o} Ceti; 14 transitions of $\rm{SO_2}$, 2 of $\rm{^{34}SO_2}$, and 1 transition each of $\rm{SO}$ and $\rm{^{34}SO}$ toward R Dor; 10 transitions of $\rm{SO_2}$, 7 of $\rm{^{34}SO_2}$, and 1 transition each of $\rm{SO}$ and $\rm{^{34}SO}$ toward R Leo; 13 transitions of $\rm{SO_2}$ and 3 of $\rm{^{34}SO_2}$ toward W Hya; and 5 transitions of $\rm{SO_2}$ and 1 of $\rm{^{34}SO_2}$ toward EP Aqr.

\item We estimate the excitation temperature using detected sulfur-bearing species to be around 200–600 K across different sources and report the column densities  $\rm{\sim(1-7)\times10^{16}\ cm^{-2}}$ of SO$_2$ and $\rm{\sim(0.8-3)\times10^{15}\ cm^{-2}}$ for $\rm{^{34}SO_2}$. Based on the spatial distribution analysis and temperature measurement, we conclude that the detected molecule traces the intermediate region ($\sim$ 50-500 au away from the central star) of the CSE.

\item {We measure the $\rm{^{32}S}$/$\rm{^{34}S}$ ratios for five sources, and two of them, R Dor and W Hya, are found to be close to the solar value within uncertainties. For {\it o} Ceti, it is slightly higher, and for R Leo and Ep Aqr, the ratios are slightly lower. 
Overall, 13 Galactic AGB stars have measured $\rm{^{32}S}$/$\rm{^{34}S}$ ratios (see Fig. \ref{fig:32S-34S-ratio}), of which 8 are consistent with the solar value within the uncertainties, while 5 show some variations.
It is believed that nucleosynthesis in AGB stars does not significantly alter the sulfur isotopic ratio, and the observed values are expected to reflect the abundances of their natal clouds.
Therefore, the observed discrepancies in the $\rm{^{32}S}$/$\rm{^{34}S}$
ratios among Galactic AGB stars may arise from assumptions of solar metallicity, which might not hold universally, and/or variations in the excitation conditions within their CSEs.
Further analysis using different sulfur-bearing species are needed to provide deeper insight, which is beyond the scope of this work.}

\item {We find that SO$_2$, $\rm{^{34}SO_2}$, SO and  $\rm{^{34}SO}$ emission contours are slightly irregular and the emission peaks are offset from the continuum peak of {\it o} Ceti. We report an extended emission structure toward south-west, offset by $\rm{\sim 3''}$ ($\sim$ 300 au) from the continuum peak, where emissions of SO and SO$_2$ correlate with SiO molecular distribution \citep{Wong2016}. This region exhibits several transitions of SO and SO$_2$ and some other species reported in Table \ref{tab:lines-oceti-arc}, likely enhanced by shock chemistry and most probably a low-velocity shock. We estimate the temperature of the arc-like emission $\sim 133\pm20$ K.}

\item{SO and $\rm{^{34}SO_2}$ show clumpy structures in R Dor, with the peak of the molecular emission centered on the continuum peak. Our results are consistent with a previous study by \cite{Danilovich2020}.}

\item{Emission of SO$_2$,  $\rm{^{34}SO_2}$, SO and  $\rm{^{34}SO}$ transitions in R Leo appear as ordered elliptical concentric emission contours and molecular emission peak slightly offset from the continuum peak. Similar to {\it o} Ceti, low-excitation transitions in R Leo show a slightly extended feature in moment 0 maps compared to higher energy transitions.}

\item{Distribution of SO$_2$ and $\rm{^{34}SO_2}$ in W Hya appears as ordered approximately circular concentric emission contours, and where the emission peak is slightly offset from the continuum peak as {\it o} Ceti. Furthermore, we find an additional structure of strong SO$_2$ emission located in the north-west position, which is $\sim$ 10$^{''}$ ($\sim 1040$ au) offset from the central continuum peak.}

\item {For EP Aqr, we have only low resolution data with which SO$_2$ and $\rm{^{34}SO_2}$ are mostly unresolved. Hence, we can not provide detailed insight about the distribution of detected species.}

{In summary, we study five oxygen-rich AGB stars with comparable low mass-loss rates and find some differences in the distribution and morphology of the emissions from SO$_2$, $\rm{^{34}SO_2}$, SO, and $\rm{^{34}SO}$ across different sources. These differences are likely due to the chemistry being highly sensitive to the physical conditions of the sources, such as the density and temperature structures of the CSEs, source multiplicity, outflows, rotation, and any other physical processes associated with the sources. Nonetheless, the mostly centralized distributions of SO and SO$_2$ in our sample support previous results for low mass-loss rate AGB stars.} 
\end{itemize}

\section*{Data Availability}
The data underlying this article (as presented in the Appendix) are available on Zenodo under a Creative Commons Attribution license at - \url{10.5281/zenodo.17302961}

\begin{acknowledgements}
    This paper makes use of the following ALMA data: ADS/JAO.ALMA$\#$2018.1.01440.S, $\#$2018.1.00649.S, \#2018.1.00749.S, $\#$2018.1.00649.S,$\#$2016.1.01202.S, $\#$2017.A.00012.S, $\#$2019.1.00801.S. ALMA is a partnership of ESO (representing its member states), NSF (USA) and NINS (Japan), together with NRC (Canada), MOST and ASIAA (Taiwan), and KASI (Republic of Korea), in cooperation with the Republic of Chile. The Joint ALMA Observatory is operated by ESO, AUI/NRAO, and NAOJ. P.G. and M.S. acknowledge the ESGC project (project No. 335497) funded by the Research Council of Norway. TD is partly supported by the Australian Research Council through a Discovery Early Career Researcher Award (DE230100183).
\end{acknowledgements}

\bibliographystyle{aa}
\bibliography{references}{}

\begin{thebibliography}{53}
\expandafter\ifx\csname natexlab\endcsname\relax\def\natexlab#1{#1}\fi

\bibitem[{{Anders} \& {Grevesse}(1989)}]{Anders1989}
{Anders}, E. \& {Grevesse}, N. 1989, \gca, 53, 197

\bibitem[{{Asplund} {et~al.}(2021){Asplund}, {Amarsi}, \& {Grevesse}}]{Asplund21}
{Asplund}, M., {Amarsi}, A.~M., \& {Grevesse}, N. 2021, \aap, 653, A141

\bibitem[{{Beck} {et~al.}(2022){Beck}, {Lebouteiller}, {Madden}, {Iserlohe}, {Krabbe}, {Ramambason}, {Fischer}, {Ka{\'z}mierczak-Barthel}, {Latzko}, \& {P{\'e}rez-Beaupuits}}]{Beck2022}
{Beck}, A., {Lebouteiller}, V., {Madden}, S.~C., {et~al.} 2022, \aap, 665, A85

\bibitem[{{Danilovich} {et~al.}(2016){Danilovich}, {De Beck}, {Black}, {Olofsson}, \& {Justtanont}}]{Danilovich2016}
{Danilovich}, T., {De Beck}, E., {Black}, J.~H., {Olofsson}, H., \& {Justtanont}, K. 2016, \aap, 588, A119

\bibitem[{{Danilovich} {et~al.}(2017{\natexlab{a}}){Danilovich}, {Lombaert}, {Decin}, {Karakas}, {Maercker}, \& {Olofsson}}]{Taissa2017}
{Danilovich}, T., {Lombaert}, R., {Decin}, L., {et~al.} 2017{\natexlab{a}}, \aap, 602, A14

\bibitem[{{Danilovich} {et~al.}(2018){Danilovich}, {Ramstedt}, {Gobrecht}, {Decin}, {De Beck}, \& {Olofsson}}]{Danilovich2018}
{Danilovich}, T., {Ramstedt}, S., {Gobrecht}, D., {et~al.} 2018, \aap, 617, A132

\bibitem[{Danilovich {et~al.}(2020)Danilovich, Richards, Decin, van~de Sande, \& Gottlieb}]{Danilovich2020}
Danilovich, T., Richards, A.~M., Decin, L., van~de Sande, M., \& Gottlieb, C.~A. 2020, An ALMA view of SO and SO2 around oxygen-rich AGB stars

\bibitem[{{Danilovich} {et~al.}(2025){Danilovich}, {Richards}, {Van de Sande}, {Gottlieb}, {Millar}, {Karakas}, {M{\"u}ller}, {Justtanont}, {Plane}, {Etoka}, {Wallstr{\"o}m}, {Decin}, {Engels}, {Groenewegen}, {Kerschbaum}, {Khouri}, {de Koter}, {Olofsson}, {Paladini}, \& {Stancliffe}}]{danilovich2025}
{Danilovich}, T., {Richards}, A.~M.~S., {Van de Sande}, M., {et~al.} 2025, \mnras, 536, 684

\bibitem[{{Danilovich} {et~al.}(2017{\natexlab{b}}){Danilovich}, {Van de Sande}, {De Beck}, {Decin}, {Olofsson}, {Ramstedt}, \& {Millar}}]{Danilovich2017}
{Danilovich}, T., {Van de Sande}, M., {De Beck}, E., {et~al.} 2017{\natexlab{b}}, \aap, 606, A124

\bibitem[{{De Nutte} {et~al.}(2017){De Nutte}, {Decin}, {Olofsson}, {Lombaert}, {de Koter}, {Karakas}, {Milam}, {Ramstedt}, {Stancliffe}, {Homan}, \& {Van de Sande}}]{DeNutte17}
{De Nutte}, R., {Decin}, L., {Olofsson}, H., {et~al.} 2017, \aap, 600, A71

\bibitem[{{Fontani} {et~al.}(2019){Fontani}, {Rivilla}, {van der Tak}, {Mininni}, {Beltr{\'a}n}, \& {Caselli}}]{Fontani2019}
{Fontani}, F., {Rivilla}, V.~M., {van der Tak}, F.~F.~S., {et~al.} 2019, \mnras, 489, 4530

\bibitem[{{Fontani} {et~al.}(2023){Fontani}, {Roueff}, {Colzi}, \& {Caselli}}]{Fontani2023}
{Fontani}, F., {Roueff}, E., {Colzi}, L., \& {Caselli}, P. 2023, \aap, 680, A58

\bibitem[{{Francis} {et~al.}(2020){Francis}, {Johnstone}, {Herczeg}, {Hunter}, \& {Harsono}}]{Francis20}
{Francis}, L., {Johnstone}, D., {Herczeg}, G., {Hunter}, T.~R., \& {Harsono}, D. 2020, \aj, 160, 270

\bibitem[{{Ghosh} {et~al.}(2024){Ghosh}, {Das}, {Gorai}, {Mondal}, {Furuya}, {Tanaka}, \& {Shimonishi}}]{Ghosh2024}
{Ghosh}, R., {Das}, A., {Gorai}, P., {et~al.} 2024, Frontiers in Astronomy and Space Sciences, 11, 1427048

\bibitem[{{Goldsmith} \& {Langer}(1999)}]{gold99}
{Goldsmith}, P.~F. \& {Langer}, W.~D. 1999, \apj, 517, 209

\bibitem[{{Gong} {et~al.}(2023){Gong}, {Henkel}, {Menten}, {Chen}, {Zhang}, {Yan}, {Weiss}, {Langer}, {Wang}, {Mao}, {Tang}, {Yang}, {Ao}, \& {Wang}}]{Gong2023}
{Gong}, Y., {Henkel}, C., {Menten}, K.~M., {et~al.} 2023, \aap, 679, L6

\bibitem[{{Hinkle} {et~al.}(2016){Hinkle}, {Lebzelter}, \& {Straniero}}]{Hinkle16}
{Hinkle}, K.~H., {Lebzelter}, T., \& {Straniero}, O. 2016, \apj, 825, 38

\bibitem[{{Hoai} {et~al.}(2023){Hoai}, {Nhung}, {Tan}, {Darriulat}, {Diep}, {Ngoc}, {Thai}, \& {Tuan-Anh}}]{Hoai2023}
{Hoai}, D.~T., {Nhung}, P.~T., {Tan}, M.~N., {et~al.} 2023, \mnras, 518, 2034

\bibitem[{{Hoai} {et~al.}(2019){Hoai}, {Nhung}, {Tuan-Anh}, {Darriulat}, {Diep}, {Le Bertre}, {Phuong}, {Thai}, \& {Winters}}]{Hoai2019}
{Hoai}, D.~T., {Nhung}, P.~T., {Tuan-Anh}, P., {et~al.} 2019, \mnras, 484, 1865

\bibitem[{{Hoai} {et~al.}(2022){Hoai}, {Tuyet Nhung}, {Darriulat}, {Diep}, {Bich Ngoc}, {Thai}, \& {Tuan-Anh}}]{Hoai2022}
{Hoai}, D.~T., {Tuyet Nhung}, P., {Darriulat}, P., {et~al.} 2022, Vietnam Journal of Science, 64, 16

\bibitem[{{Kami{\'n}ski} {et~al.}(2017){Kami{\'n}ski}, {M{\"u}ller}, {Schmidt}, {Cherchneff}, {Wong}, {Br{\"u}nken}, {Menten}, {Winters}, {Gottlieb}, \& {Patel}}]{Kaminski2017}
{Kami{\'n}ski}, T., {M{\"u}ller}, H.~S.~P., {Schmidt}, M.~R., {et~al.} 2017, \aap, 599, A59

\bibitem[{{Khouri} {et~al.}(2024){Khouri}, {Olofsson}, {Vlemmings}, {Schirmer}, {Tafoya}, {Maercker}, {De Beck}, {Nyman}, \& {Saberi}}]{khouri2024}
{Khouri}, T., {Olofsson}, H., {Vlemmings}, W.~H.~T., {et~al.} 2024, \aap, 685, A11

\bibitem[{{Khouri} {et~al.}(2018){Khouri}, {Vlemmings}, {Olofsson}, {Ginski}, {De Beck}, {Maercker}, \& {Ramstedt}}]{Khouri18}
{Khouri}, T., {Vlemmings}, W.~H.~T., {Olofsson}, H., {et~al.} 2018, \aap, 620, A75

\bibitem[{{Krijt} {et~al.}(2023){Krijt}, {Kama}, {McClure}, {Teske}, {Bergin}, {Shorttle}, {Walsh}, \& {Raymond}}]{Krijt2023}
{Krijt}, S., {Kama}, M., {McClure}, M., {et~al.} 2023, in Astronomical Society of the Pacific Conference Series, Vol. 534, Protostars and Planets VII, ed. S.~{Inutsuka}, Y.~{Aikawa}, T.~{Muto}, K.~{Tomida}, \& M.~{Tamura}, 1031

\bibitem[{{Lindqvist} {et~al.}(1988){Lindqvist}, {Nyman}, {Olofsson}, \& {Winnberg}}]{Lindqvist1988}
{Lindqvist}, M., {Nyman}, L.~A., {Olofsson}, H., \& {Winnberg}, A. 1988, \aap, 205, L15

\bibitem[{{Mart{\'\i}n} {et~al.}(2021){Mart{\'\i}n}, {Mangum}, {Harada}, {Costagliola}, {Sakamoto}, {Muller}, {Aladro}, {Tanaka}, {Yoshimura}, {Nakanishi}, {Herrero-Illana}, {M{\"u}hle}, {Aalto}, {Behrens}, {Colzi}, {Emig}, {Fuller}, {Garc{\'\i}a-Burillo}, {Greve}, {Henkel}, {Holdship}, {Humire}, {Hunt}, {Izumi}, {Kohno}, {K{\"o}nig}, {Meier}, {Nakajima}, {Nishimura}, {Padovani}, {Rivilla}, {Takano}, {van der Werf}, {Viti}, \& {Yan}}]{Martin2021}
{Mart{\'\i}n}, S., {Mangum}, J.~G., {Harada}, N., {et~al.} 2021, \aap, 656, A46

\bibitem[{Massalkhi {et~al.}(2020)Massalkhi, Agúndez, Cernicharo, \& Velilla-Prieto}]{Massalkhi2020}
Massalkhi, S., Agúndez, M., Cernicharo, J., \& Velilla-Prieto, L. 2020, Astronomy and Astrophysics, 641

\bibitem[{{Mauersberger} {et~al.}(2004){Mauersberger}, {Ott}, {Henkel}, {Cernicharo}, \& {Gallino}}]{Mauersberger2004}
{Mauersberger}, R., {Ott}, U., {Henkel}, C., {Cernicharo}, J., \& {Gallino}, R. 2004, \aap, 426, 219

\bibitem[{{McMullin} {et~al.}(2007){McMullin}, {Waters}, {Schiebel}, {Young}, \& {Golap}}]{mcmu07}
{McMullin}, J.~P., {Waters}, B., {Schiebel}, D., {Young}, W., \& {Golap}, K. 2007, in Astronomical Society of the Pacific Conference Series, Vol. 376, Astronomical Data Analysis Software and Systems XVI, ed. R.~A. {Shaw}, F.~{Hill}, \& D.~J. {Bell}, 127

\bibitem[{{Mifsud} {et~al.}(2021){Mifsud}, {Ka{\r{A}}uchov{\'a}}, {Herczku}, {Ioppolo}, {Juh{\'a}sz}, {Kov{\'a}cs}, {Mason}, {McCullough}, \& {Sulik}}]{Mifsud2021}
{Mifsud}, D.~V., {Ka{\r{A}}uchov{\'a}}, Z., {Herczku}, P., {et~al.} 2021, \ssr, 217, 14

\bibitem[{{Mouhcine} {et~al.}(2005){Mouhcine}, {Rich}, {Ferguson}, {Brown}, \& {Smith}}]{Mouhcine2005}
{Mouhcine}, M., {Rich}, R.~M., {Ferguson}, H.~C., {Brown}, T.~M., \& {Smith}, T.~E. 2005, \apj, 633, 828

\bibitem[{{M{\"u}ller} {et~al.}(2005){M{\"u}ller}, {Schl{\"o}der}, {Stutzki}, \& {Winnewisser}}]{mull05}
{M{\"u}ller}, H. S.~P., {Schl{\"o}der}, F., {Stutzki}, J., \& {Winnewisser}, G. 2005, Journal of Molecular Structure, 742, 215

\bibitem[{{M{\"u}ller} {et~al.}(2001){M{\"u}ller}, {Thorwirth}, {Roth}, \& {Winnewisser}}]{mull01}
{M{\"u}ller}, H.~S.~P., {Thorwirth}, S., {Roth}, D.~A., \& {Winnewisser}, G. 2001, \aap, 370, L49

\bibitem[{{Nhung} {et~al.}(2015){Nhung}, {Hoai}, {Winters}, {Le Bertre}, {Diep}, {Phuong}, {Thao}, {Tuan-Anh}, \& {Darriulat}}]{Nhung2015}
{Nhung}, P.~T., {Hoai}, D.~T., {Winters}, J.~M., {et~al.} 2015, \aap, 583, A64

\bibitem[{{Omont} {et~al.}(1993){Omont}, {Lucas}, {Morris}, \& {Guilloteau}}]{Omont93}
{Omont}, A., {Lucas}, R., {Morris}, M., \& {Guilloteau}, S. 1993, \aap, 267, 490

\bibitem[{{Saberi} {et~al.}(2022){Saberi}, {Khouri}, {Velilla-Prieto}, {Fonfr{\'\i}a}, {Vlemmings}, \& {Wedemeyer}}]{saberi2022}
{Saberi}, M., {Khouri}, T., {Velilla-Prieto}, L., {et~al.} 2022, \aap, 663, A54

\bibitem[{{Shimonishi} {et~al.}(2020){Shimonishi}, {Das}, {Sakai}, {Tanaka}, {Aikawa}, {Onaka}, {Watanabe}, \& {Nishimura}}]{Shimonishi2020}
{Shimonishi}, T., {Das}, A., {Sakai}, N., {et~al.} 2020, \apj, 891, 164

\bibitem[{{Shimonishi} {et~al.}(2021){Shimonishi}, {Izumi}, {Furuya}, \& {Yasui}}]{Shimonishi2021}
{Shimonishi}, T., {Izumi}, N., {Furuya}, K., \& {Yasui}, C. 2021, \apj, 922, 206

\bibitem[{{Stanghellini} {et~al.}(2015){Stanghellini}, {Magrini}, \& {Casasola}}]{Stanghellini2015}
{Stanghellini}, L., {Magrini}, L., \& {Casasola}, V. 2015, \apj, 812, 39

\bibitem[{{Suutarinen} {et~al.}(2014){Suutarinen}, {Kristensen}, {Mottram}, {Fraser}, \& {van Dishoeck}}]{Suutarinen2014}
{Suutarinen}, A.~N., {Kristensen}, L.~E., {Mottram}, J.~C., {Fraser}, H.~J., \& {van Dishoeck}, E.~F. 2014, \mnras, 440, 1844

\bibitem[{{Tuan-Anh} {et~al.}(2019){Tuan-Anh}, {Hoai}, {Nhung}, {Darriulat}, {Diep}, {Le Bertre}, {Phuong}, {Thai}, \& {Winters}}]{Tuan2019}
{Tuan-Anh}, P., {Hoai}, D.~T., {Nhung}, P.~T., {et~al.} 2019, \mnras, 487, 622

\bibitem[{{Unnikrishnan} {et~al.}(2024){Unnikrishnan}, {De Beck}, {Nyman}, {Olofsson}, {Vlemmings}, {Tafoya}, {Maercker}, {Charnley}, {Cordiner}, {de Gregorio}, {Humphreys}, {Millar}, \& {Rawlings}}]{Unnikrishnan2024}
{Unnikrishnan}, R., {De Beck}, E., {Nyman}, L.~{\r{A}}., {et~al.} 2024, \aap, 684, A4

\bibitem[{{Vastel} {et~al.}(2015){Vastel}, {Bottinelli}, {Caux}, {Glorian}, \& {Boiziot}}]{vast15}
{Vastel}, C., {Bottinelli}, S., {Caux}, E., {Glorian}, J.~M., \& {Boiziot}, M. 2015, in SF2A-2015: Proceedings of the Annual meeting of the French Society of Astronomy and Astrophysics, 313--316

\bibitem[{{Velilla Prieto} {et~al.}(2017){Velilla Prieto}, {S{\'a}nchez Contreras}, {Cernicharo}, {Ag{\'u}ndez}, {Quintana-Lacaci}, {Bujarrabal}, {Alcolea}, {Balan{\c{c}}a}, {Herpin}, {Menten}, \& {Wyrowski}}]{VelillaPrieto17}
{Velilla Prieto}, L., {S{\'a}nchez Contreras}, C., {Cernicharo}, J., {et~al.} 2017, \aap, 597, A25

\bibitem[{{Vlemmings} {et~al.}(2019){Vlemmings}, {Khouri}, \& {Olofsson}}]{Vlemmings19}
{Vlemmings}, W.~H.~T., {Khouri}, T., \& {Olofsson}, H. 2019, \aap, 626, A81

\bibitem[{{Vlemmings} {et~al.}(2015){Vlemmings}, {Ramstedt}, {O'Gorman}, {Humphreys}, {Wittkowski}, {Baudry}, \& {Karovska}}]{Vlemmings2015}
{Vlemmings}, W.~H.~T., {Ramstedt}, S., {O'Gorman}, E., {et~al.} 2015, \aap, 577, L4

\bibitem[{{Wallstr{\"o}m} {et~al.}(2024){Wallstr{\"o}m}, {Danilovich}, {M{\"u}ller}, {Gottlieb}, {Maes}, {Van de Sande}, {Decin}, {Richards}, {Baudry}, {Bolte}, {Ceulemans}, {De Ceuster}, {de Koter}, {El Mellah}, {Esseldeurs}, {Etoka}, {Gobrecht}, {Gottlieb}, {Gray}, {Herpin}, {Jeste}, {Kee}, {Kervella}, {Khouri}, {Lagadec}, {Malfait}, {Marinho}, {McDonald}, {Menten}, {Millar}, {Montarg{\`e}s}, {Nuth}, {Plane}, {Sahai}, {Waters}, {Wong}, {Yates}, \& {Zijlstra}}]{Wallstrom2024}
{Wallstr{\"o}m}, S.~H.~J., {Danilovich}, T., {M{\"u}ller}, H.~S.~P., {et~al.} 2024, \aap, 681, A50

\bibitem[{{Wang} {et~al.}(2004){Wang}, {Henkel}, {Chin}, {Whiteoak}, {Hunt Cunningham}, {Mauersberger}, \& {Muders}}]{Wang2004}
{Wang}, M., {Henkel}, C., {Chin}, Y.~N., {et~al.} 2004, \aap, 422, 883

\bibitem[{{Westerlund}(1997)}]{Westerlund1997}
{Westerlund}, B.~E. 1997, Cambridge Astrophysics Series, 29

\bibitem[{Wong {et~al.}(2016)Wong, Kamiński, Menten, \& Wyrowski}]{Wong2016}
Wong, K.~T., Kamiński, T., Menten, K.~M., \& Wyrowski, F. 2016, Astronomy and Astrophysics, 590

\bibitem[{{Woosley} \& {Weaver}(1995)}]{Woosley95}
{Woosley}, S.~E. \& {Weaver}, T.~A. 1995, \apjs, 101, 181

\bibitem[{{Yamamura} {et~al.}(1999){Yamamura}, {de Jong}, {Onaka}, {Cami}, \& {Waters}}]{Yamamura1999}
{Yamamura}, I., {de Jong}, T., {Onaka}, T., {Cami}, J., \& {Waters}, L.~B.~F.~M. 1999, \aap, 341, L9

\bibitem[{{Yan} {et~al.}(2023){Yan}, {Henkel}, {Kobayashi}, {Menten}, {Gong}, {Zhang}, {Yu}, {Yang}, {Xie}, \& {Wang}}]{Yan2023}
{Yan}, Y.~T., {Henkel}, C., {Kobayashi}, C., {et~al.} 2023, \aap, 670, A98

\end{thebibliography}

\begin{appendix}
\onecolumn

\section{Summary of detected species and their transitions}
In this section we present the line parameters of all detected species towards a sample of five AGB stars and summary of detected transitions towards {\it o} Ceti arc-like structure
\footnotesize
\begin{longtable}{|c|c|c|c|c|c|c|c|c|c|}
\caption{Line parameters of observed transitions towards our sample AGB stars. \label{tab:lines}}\\
\hline
\cline{1-10}
Species & QNs & Freq & log$_{10}(\rm{A_{ij}})$ & E$_{up}$ & $\rm{S_{ij}\mu^{2}}$ & Peak Flux & FWHM & Integrated Intensity & $\tau$ \\
        &     & (MHz) & (s$^{-1}$) & (K) & (D$^2$) & (K) & (km s$^{-1}$) & (K km s$^{-1}$)&\\
\endfirsthead
\caption{continued}\\
\hline
Species & QNs & Freq & log$_{10}(\rm{A_{ij}})$ & E$_{up}$ & $\rm{S_{ij}\mu^{2}}$ & Peak Flux & FWHM & Integrated Intensity & $\tau$ \\
        &     & (MHz) & (s$^{-1}$) & (K) & (D$^2$) & (K) & (km s$^{-1}$) & (K km s$^{-1}$) & \\
\hline
\endhead
\hline
\endfoot
\hline
\endlastfoot
\hline
\multicolumn{10}{|c|}{{\it o} Ceti} \\
\hline
\cline{1-10}
SO$_2$  & 7(4,4) - 6(3,3) & 491934.72& -3.02 & 65.01 & 10.27 & 13.89 $\pm$ 0.54 & 8.63 $\pm$ 0.39 & 126.97 $\pm$ 7.56& 0.0265\\ 
SO$_2$  & 12(3,9) - 11(2,10) & 494779.73 & -3.26 & 93.96 & 9.80 & 12.24 $\pm$ 0.37 & 9.12 $\pm$ 0.32 & 118.34 $\pm$ 5.53& 0.0234 \\ 
SO$_2$  & 25(2,24) - 24(1,23) & 482503.15 & -2.98 & 303.69 & 40.61 & 22.89 $\pm$ 0.23 & 8.79 $\pm$ 0.10 & 213.32 $\pm$ 3.30& 0.0441\\ 
SO$_2$  & 19(9,11) - 20(8,12) & 481166.24 & -3.83 & 372.91 & 4.44 & 3.15 $\pm$ 0.29 & 9.64 $\pm$ 1.03 & 32.16 $\pm$ 4.52& 0.0060\\ 
SO$_2$ & 38(2,36) - 39(1,39) & 481237.15 & -5.39 & 697.69 & 0.24 & 1.80 $\pm$ 0.13 & 8.03 $\pm$ 0.67 & 15.34 $\pm$ 1.68& 0.0034\\ 
SO$_2$ & 11(6,6) - 12(5, 7) & 331580.24 & -4.36 & 148.95 & 2.36 & 2.25 $\pm$ 0.20 & 8.41 $\pm$ 0.85 & 20.04 $\pm$ 2.69& 0.0042\\ 
SO$_2$ & 21(2,20) - 21(1,21) & 332091.43 & -3.82 & 219.53 & 15.20 & 8.74 $\pm$ 0.23 & 9.74 $\pm$ 0.30 & 90.20 $\pm$ 3.64& 0.0165\\ 
SO$_2$ & 4(3,1) - 3(2, 2) & 332505.24 & -3.48 & 31.29 & 6.92 & 7.00 $\pm$ 0.18 & 8.89 $\pm$ 0.27 & 65.94 $\pm$ 2.62& 0.0132\\ 
SO$_2$ & 13(2,12) - 12(1,11) & 345338.54 & -3.62 & 92.98 & 13.41 & 12.50 $\pm$ 0.23 & 8.47 $\pm$ 0.18 & 112.19 $\pm$ 3.14& 0.0237\\ 
SO$_2$ & 20(1,19) - 20(0,20) & 282292.81 & -4.00 & 198.88 & 15.69 & 10.17 $\pm$ 0.25 & 8.97 $\pm$ 0.26 & 96.77 $\pm$ 3.67& 0.0192\\ 
SO$_2$ & 24(8,16) - 25(7,19) & 282636.22 & -4.36 & 432.62 & 8.12 & 3.66 $\pm$ 0.15 & 8.40 $\pm$ 0.39 & 32.61 $\pm$ 1.99& 0.0069\\ 
SO$_2$ & 16(0,16) - 15(1,15) & 283464.77 & -3.57 & 121.03 & 33.60 & 18.98 $\pm$ 0.34 & 8.93 $\pm$ 0.18 & 179.68 $\pm$ 4.91& 0.0361\\ 
SO$_2$ & 17(3,15) - 17(2,16) & 285743.59 & -3.74 & 162.93 & 23.41 & 13.45 $\pm$ 0.21 & 8.97 $\pm$ 0.16 & 127.86 $\pm$ 3.02& 0.0255\\ 
SO$_2$ & 26(2,24) - 26(1,25) & 296168.67 & -3.73 & 340.55 & 32.84 & 10.67 $\pm$ 0.22 & 8.44 $\pm$ 0.20 & 95.43 $\pm$ 3.05& 0.0202\\ 
SO$_2$ & 24(4,20) - 24(3,21) & 296535.42 & -3.59 & 316.58 & 41.34 & 38.12 $\pm$ 2.47 & 8.40 $\pm$ 0.63 & 339.51 $\pm$ 33.53& 0.0739\\ 
SO$_2$ & 18(7,11) - 19(6,14) & 297782.59 & -4.34 & 277.30 & 5.48 & 3.32 $\pm$ 0.21 & 8.66 $\pm$ 0.64 & 30.49 $\pm$ 2.96& 0.0062\\ 
$\rm{^{34}SO_2}$ & 7(4,4) - 6(3,3) & 479309.99 & -3.06 & 63.70 & 10.25 & 1.31 $\pm$ 0.08 & 11.14 $\pm$ 0.96 & 15.46 $\pm$ 1.62& 0.0025\\ 
$\rm{^{34}SO_2}$ & 27(1,27) - 26(0,26) & 482025.14 & -2.82 & 329.28 & 63.92 & 3.14 $\pm$ 0.14 & 9.77 $\pm$ 0.56 & 32.47 $\pm$ 2.34& 0.0059\\ 
$\rm{^{34}SO_2}$ & 5(4,2) - 5(3,3) & 345651.29 & -3.71 & 51.76 & 4.49 & 0.77 $\pm$ 0.12 & 6.80 $\pm$ 1.18 & 5.53 $\pm$ 1.27& 0.0014\\ 
$\rm{^{34}SO_2}$ & 4(4,0) - 4(3,1) & 345678.79 & -3.88 & 47.17 & 2.44 & 0.77 $\pm$ 0.12 & 6.80 $\pm$ 1.24 & 5.53 $\pm$ 1.34& 0.0014\\ 
$\rm{^{34}SO_2}$ & 6(4,2) - 6(3,3) & 345553.09 & -3.63 & 57.27 & 6.34 & 0.60 $\pm$ 0.29 & 5.98 $\pm$ 3.34 & 3.80 $\pm$ 2.80& 0.0011\\ 
$\rm{^{34}SO_2}$ & 7(4,4) - 7(3,5) & 345519.66 & -3.59 & 63.70 & 8.09 & 1.12 $\pm$ 0.14 & 8.49 $\pm$ 1.26 & 10.10 $\pm$ 1.98& 0.0021\\ 
$\rm{^{34}SO_2}$ & 8(4,4) - 8(3,5) & 345168.66 & -3.56 & 71.05 & 9.77 & 0.78 $\pm$ 0.14 & 5.57 $\pm$ 1.18 & 4.58 $\pm$ 1.28& 0.0015\\ 
$\rm{^{34}SO_2}$ & 9(4,6) - 9(3,7) & 345285.62 & -3.54 & 79.32 & 11.41 & 0.92 $\pm$ 0.09 & 8.14 $\pm$ 0.97 & 7.93 $\pm$ 1.25& 0.0017 \\ 
$\rm{^{34}SO_2}$ & 13(4,10) - 13(3,11) & 344807.91 & -3.50 & 121.63 & 17.92 & 0.85 $\pm$ 0.67 & 9.70 $\pm$ 8.77 & 8.78 $\pm$ 10.51& 0.0016\\ 
$\rm{^{34}SO_2}$ & 16(4,12) - 16(3,13) & 332836.22 & -3.52 & 163.13 & 23.23 & 1.10 $\pm$ 0.15 & 8.82 $\pm$ 1.39 & 10.31 $\pm$ 2.15 &  0.0021 \\ 
$\rm{^{34}SO_2}$ & 19(1,19) - 18(0,18) & 344581.04 & -3.29 & 167.66 & 42.24 & 4.16 $\pm$ 0.22 & 8.28 $\pm$ 0.50 & 36.54 $\pm$ 2.92 & 0.0078\\
$\rm{^{34}SO_2}$ & 17(4,14) - 17(3,15) & 345929.35 & -3.47 & 178.77 & 24.49 & 1.09 $\pm$ 0.13 & 8.26 $\pm$ 1.16 & 9.54 $\pm$ 1.77 & 0.0020\\
$\rm{^{34}SO_2}$ & 15(7,9) - 16(6,10) & 333683.92 & -4.28 & 226.38 & 3.78 & 1.38 $\pm$ 0.17 & 5.64 $\pm$ 0.78 & 8.24 $\pm$ 1.51 & 0.0026\\ 
$\rm{^{34}SO_2}$ & 28(3,25) - 27(4,24) & 333364.11 & -3.93 & 402.63 & 15.68 & 0.60 $\pm$ 0.10 & 11.66 $\pm$ 2.25 & 7.45 $\pm$ 1.90 & 0.0011  \\ 
SO & 10(9) - 9(9) & 295355.70 & -5.29 & 120.24 & 0.33 & 4.42 $\pm$ 0.17 & 5.92 $\pm$ 0.27 & 27.75 $\pm$ 1.64 & 0.0083 \\ 
SO & 7(6) - 6(5) & 296550.06& -3.49 & 64.89 & 13.83 & 38.12 $\pm$ 11.86 & 8.40 $\pm$ 3.02 & 339.48 $\pm$ 161.25 & 0.0739 \\ 
SO & 8(8) - 7(7) & 344310.61& -3.29 & 87.48 & 18.56 & 41.14 $\pm$ 0.49 & 8.19 $\pm$ 0.11 & 356.97 $\pm$ 6.47 & 0.0802\\ 
$\rm{^{34}SO}$ & 8(7) - 7(6) & 333900.98 & -3.32 &79.86 &16.24  & 7.00 $\pm$ 0.26 & 6.66 $\pm$ 0.29 & 46.62 $\pm$ 7.24 &  0.0132\\ 
$\rm{^{34}SO}$ & 7(7) - 6(6) & 295396.33 & -3.49 &69.87 & 16.15 & 8.83 $\pm$ 0.23 & 6.17 $\pm$ 0.19 & 57.74 $\pm$ 3.72 &  0.0166\\ 
\cline{1-10}
\multicolumn{10}{|c|}{R Dor}\\
\cline{1-10}
SO$_2$ & 10(3,7) - 10(0,10) & 479317.52 & -4.92 & 72.71 & 0.20 & 0.09 $\pm$ 0.01 & 12.52 $\pm$ 2.01 & 1.18 $\pm$ 0.25 & 0.0002 \\ 
SO$_2$& 19(9,11) - 20(8,12) & 481166.24 & -3.83 & 372.91 & 4.44 & 0.11 $\pm$ 0.02 & 8.25 $\pm$ 1.84 & 0.98 $\pm$ 0.29 & 0.0003\\ 
SO$_2$ & 25(2,24) - 24(1,23) & 482503.15 & -2.98 & 303.69 & 40.61 & 2.42 $\pm$ 0.07 & 7.73 $\pm$ 0.24 & 19.85 $\pm$ 0.82 & 0.0062 \\ 
SO$_2$ & 7(4,4) - 6(3,3) & 491934.72 & -3.02 & 65.01 & 10.27 & 1.50 $\pm$ 0.04 & 7.75 $\pm$ 0.24 & 12.28 $\pm$ 0.50 & 0.0038\\ 
SO$_2$ & 12(3,9) - 11(2,10) & 494779.73 & -3.26 & 93.96 & 9.80 & 1.49 $\pm$ 0.06 & 8.04 $\pm$ 0.36 & 12.68 $\pm$ 0.75 & 0.0038\\ 
SO$_2$ & 18(0,18) - 17(1,17) & 321330.17 & -3.39 & 151.50 & 39.15 & 50.06 $\pm$ 1.36 & 9.53 $\pm$ 0.30 & 505.98 $\pm$ 20.99 & 0.1345 \\ 
SO$_2$ & 40(5,35) - 39(6,34) & 321420.53 & -4.08 & 823.69 & 17.34 & 6.45 $\pm$ 0.58 & 7.53 $\pm$ 0.78 & 51.44 $\pm$ 7.02 & 0.0162\\ 
SO$_2$ & 22(8,14) - 23(7,17) & 321782.62 & -4.22 & 389.37 & 6.95 & 4.89 $\pm$ 0.85 & 4.49 $\pm$ 0.89 & 23.25 $\pm$ 6.13 & 0.0122 \\ 
SO$_2$ & 34(4,30) - 33(5,29) & 322475.41 & -4.05 & 594.66 & 15.82 & 10.28 $\pm$ 0.72 & 9.89 $\pm$ 0.80 & 107.77 $\pm$ 11.55 & 0.0261 \\ 
SO$_2$ & 11(2,10) - 10(1,9) & 323026.46 & -3.73 & 70.21 & 10.90 & 21.89 $\pm$ 1.07 & 9.40 $\pm$ 0.53 & 218.00 $\pm$ 16.28 & 0.0565 \\ 
SO$_2$ & 22(1,21) - 22(0,22) & 323526.42 & -3.87 & 238.04 & 15.57 & 21.47 $\pm$ 1.50 & 9.87 $\pm$ 0.80 & 224.69 $\pm$ 23.93 & 0.0554 \\ 
SO$_2$ & 8(2,6) - 7(1,7) & 334673.35 & -3.90 & 43.15 & 4.95 & 10.13 $\pm$ 0.38 & 9.93 $\pm$ 0.43 & 106.65 $\pm$ 6.18 &  0.0257\\ 
SO$_2$ & 23(3,21) - 23(2,22) & 336089.23 & -3.57 & 276.02 & 28.41 & 30.17 $\pm$ 0.56 & 8.70 $\pm$ 0.19 & 278.10 $\pm$ 7.88 & 0.0789 \\ 
SO$_2$ & 16( 7, 9) - 17(6,12) & 336669.58 & -4.23 & 245.11 & 4.34 & 5.57 $\pm$ 0.39 & 7.56 $\pm$ 0.61 & 44.68 $\pm$ 4.80 & 0.0140 \\ 
$\rm{^{34}SO_2}$ & 18(4,14) - 18(3,15) & 323806.69 & -3.54 & 195.47 & 27.04 & 4.11 $\pm$ 2.12 & 5.16 $\pm$ 3.09 & 22.48 $\pm$ 17.77 & 0.0130 \\ 
$\rm{^{34}SO_2}$ & 27(1,27) - 26(0,26) & 482025.14 & -2.82 & 329.28 & 63.92 & 0.23 $\pm$ 0.04 & 6.12 $\pm$ 1.06 & 1.52 $\pm$ 0.35 & 0.0006\\ 
SO & 11(10) - 10(10) & 336553.80 & -5.21 & 142.88 & 4.34 & 0.92 $\pm$ 0.12 & 6.61 $\pm$ 0.97 & 6.44 $\pm$ 3.1 & 0.0022 \\ 
$\rm{^{34}SO}$ & 8(7) - 7(6) & 333900.98 & -3.32 & 79.86 & 16.24 & 2.95 $\pm$ 0.2616 & 9.60 $\pm$ 1.12 & 30.02 $\pm$ 6.30 & 0.0073 \\ 
\cline{1-10}
\multicolumn{10}{|c|}{R Leo}\\
\cline{1-10}
SO$_2$ & 25(2,24) - 24(1,23) & 482503.15 & -2.98 & 303.69 & 40.61 & 0.46 $\pm$ 0.02 & 7.98 $\pm$ 0.46 & 3.90 $\pm$ 0.30 & 0.0012  \\ 
SO$_2$ & 7(4, 4) - 6(3,3) & 491934.72 & -3.02 & 65.01 & 10.27 & 0.47 $\pm$ 0.03 & 6.19 $\pm$ 0.51 & 3.07 $\pm$ 0.33 & 0.0012\\ 
SO$_2$ & 12(3, 9) - 11(2,10) & 494779.73 & -3.26 & 93.96 & 9.80 & 0.26 $\pm$ 0.03 & 7.58 $\pm$ 0.99 & 2.12 $\pm$ 0.36 & 0.0007\\ 
SO$_2$ & 28(2,26) - 28(1,27) & 340316.41 & -3.59 & 391.79 & 32.05 & 4.61 $\pm$ 0.08 & 8.32 $\pm$ 0.16 & 40.67 $\pm$ 1.05 &0.0119\\ 
SO$_2$ & 21(8,14) - 22(7,15) & 341275.52 & -4.16 & 369.13 & 6.37 & 2.20 $\pm$ 0.07 & 7.30 $\pm$ 0.27 & 17.01 $\pm$ 0.83 & 0.0056\\ 
SO$_2$ & 40(4,36) - 40(3,37) & 341403.07 & -3.39 & 808.36 & 71.97 & 8.47 $\pm$ 0.14 & 6.39 $\pm$ 0.13 & 57.39 $\pm$ 1.49 &0.0221\\ 
SO$_2$ & 36(5,31) - 36(4,32) & 341673.96 & -3.36 & 678.51 & 68.31 & 6.93 $\pm$ 0.12 & 7.46 $\pm$ 0.15 & 54.83 $\pm$ 1.47 & 0.0181\\ 
SO$_2$ & 34(3,31) - 34(2,32) & 342761.63 & -3.46 & 581.92 & 50.74 & 3.77 $\pm$ 0.09 & 9.78 $\pm$ 0.26 & 39.05 $\pm$ 1.35 & 0.0097\\ 
SO$_2$ & 14(4,10)-14(3,11) & 351873.87 & -3.46 & 135.87 & 19.63 & 4.76 $\pm$ 0.10 & 8.14 $\pm$ 0.20 & 41.03 $\pm$ 1.36 & 0.0124\\ 
SO$_2$ & 12(4, 8) - 12(3,9) & 355045.52 & -3.47 & 111.00 & 16.30 & 24.23 $\pm$ 0.41 & 4.53 $\pm$ 0.09 & 116.36 $\pm$ 3.01 & 0.0650\\ 
 $\rm{^{34}SO_2}$ & 12(4,8) - 12(3,9) & 342332.01 & -3.52 & 109.67 & 16.34 & 0.62 $\pm$ 0.07 & 7.46 $\pm$ 1.03 & 4.87 $\pm$ 0.89 & 0.0015\\ 
\hline
$\rm{^{34}SO_2}$ & 14(7,7) - 15(6,10) & 353002.39 & -4.24 & 212.59 & 3.23 & 0.08 $\pm$ 0.08 & 3.69 $\pm$ 3.99 & 0.32 $\pm$ 0.45 & 0.0001\\ 
$\rm{^{34}SO_2}$ & 21(4,18) - 21(3,19) & 352082.92 & -3.44 & 250.78 & 30.92 & 0.79 $\pm$ 0.04 & 12.81 $\pm$ 0.77 & 10.67 $\pm$ 0.85 & 0.0020  \\ 
$\rm{^{34}SO_2}$ & 19(8,12) - 20(7,13) & 354397.76 & -4.16 & 326.16 & 5.23 & 0.39 $\pm$ 0.11 & 2.71 $\pm$ 0.90 & 1.11 $\pm$ 0.48 & 0.0009\\ 
$\rm{^{34}SO_2}$ & 24(9,15) - 25(8,18) & 355322.20 & -4.11 & 467.40 & 7.27 & 0.24 $\pm$ 0.11 & 2.71 $\pm$ 1.34 & 0.70 $\pm$ 0.46 & 0.0005\\
$\rm{^{34}SO_2}$ & 34(3,31) - 34(2,32) & 354277.56 & -3.44 & 581.30 & 48.47 & 0.79 $\pm$ 0.08 & 5.48 $\pm$ 0.64 & 4.61 $\pm$ 0.71 & 0.0020\\ 
$\rm{^{34}SO_2}$ & 40(4,36) - 40(3,37) & 353949.96 & -3.36 & 807.68 & 68.60 & 0.58 $\pm$ 0.08 & 6.68 $\pm$ 1.03 & 4.09 $\pm$ 0.84 & 0.0014\\ 
SO & 7(8) - 6(7) & 340714.16 & -3.30 & 81.24 & 16.24 & 18.11 $\pm$ 0.23 & 7.09 $\pm$ 0.10 & 136.05 $\pm$ 2.62 & 0.0481\\ 
$\rm{^{34}SO}$ & 8(9) - 7(8) & 339857.26 & -3.29 & 77.34 &21.11 & 6.24 $\pm$ 0.25 & 3.40 $\pm$ 0.11 & 22.48 $\pm$ 1.70 & 0.0162\\ 
\cline{1-10}
\multicolumn{10}{|c|}{W Hya}\\
\cline{1-10}
SO$_2$ & 10(3, 7) - 10(0,10) & 479317.52 & -4.92 & 72.71 & 0.20 & 0.15 $\pm$ 0.04 & 6.28 $\pm$ 1.92 & 1.02 $\pm$ 0.41 & 0.0002\\ 
SO$_2$ & 19(9,11) - 20(8,12) & 481166.24 & -3.83 & 372.91 & 4.44 & 0.18 $\pm$ 0.04 & 7.41 $\pm$ 1.77 & 1.43 $\pm$ 0.45 & 0.0003\\ 
SO$_2$& 25(2,24) - 24(1,23) & 482503.15 & -2.98 & 303.69 & 40.61 & 1.14 $\pm$ 0.04 & 9.08 $\pm$ 0.33 & 10.97 $\pm$ 0.53 & 0.0019 \\ 
SO$_2$ & 7(4,4) - 6(3,3) & 491934.72 & -3.02 & 65.01 & 10.27 & 0.74 $\pm$ 0.04 & 8.35 $\pm$ 0.54 & 6.59 $\pm$ 0.57 & 0.0012\\ 
SO$_2$ & 12(3,9) - 11(2,10) & 494779.73 & -3.26 & 93.96 & 9.80 & 0.48 $\pm$ 0.07 & 7.30 $\pm$ 1.16 & 3.74 $\pm$ 0.79 & 0.0008\\
SO$_2$ & 36(10,26) - 37(9,29) & 250816.79 & -4.47 & 857.17 & 13.40 & 1.75 $\pm$ 0.12 & 7.27 $\pm$ 0.57 & 13.48 $\pm$ 1.40 & 0.0027\\ 
SO$_2$ & 32(4,28) - 31(5,27) & 252563.89 & -4.38 & 531.10 & 14.32 & 3.04 $\pm$ 0.09 & 6.79 $\pm$ 0.24 & 21.87 $\pm$ 1.01 & 0.0048\\ 
SO$_2$ & 38(5,33) - 37(6,32) & 253935.88 & -4.39 & 749.09 & 16.29 & 3.52 $\pm$ 1.28 & 6.72 $\pm$ 2.34 & 25.05 $\pm$ 12.62 & 0.0056\\ 
SO$_2$ & 30(9,21) - 31(8,24) & 266943.32 & -4.41 & 625.92 & 10.76 & 3.84 $\pm$ 0.07 & 5.87 $\pm$ 0.12 & 23.90 $\pm$ 0.65 & 0.0062\\ 
SO$_2$ & 13(3,11) - 13(2,12) & 267537.45 & -3.82 & 105.82 & 18.32 & 8.55 $\pm$ 0.12 & 8.60 $\pm$ 0.14 & 77.99 $\pm$ 1.63 & 0.0141\\ 
SO$_2$ & 28(4,24) - 28(3,25) & 267719.84 & -3.67 & 415.88 & 55.04 & 12.96 $\pm$ 0.19 & 7.13 $\pm$ 0.12 & 97.94 $\pm$ 2.18 &  0.0215\\ 
SO$_2$ & 9(5,5) - 10(4, 6) & 268168.33 & -4.62 & 102.70 & 2.02 & 4.86 $\pm$ 6.57 & 6.34 $\pm$ 9.89 & 32.63 $\pm$ 67.40 & 0.0079\\ 
SO$_2$ & 11(3,9) - 11(2,10) & 268169.79 & -3.81 & 844.26 & 15.81 & 4.86 $\pm$ 6.62 & 6.34 $\pm$ 9.96 & 32.63 $\pm$ 67.86 & 0.0079\\ 
$\rm{^{34}SO_2}$ & 11(3,9) - 11(2,10) & 253936.32 & -3.89 & 82.05 & 15.42 & 3.52 $\pm$ 1.29 & 6.72 $\pm$ 2.36 & 25.05 $\pm$ 12.73 & 0.0056\\ 
$\rm{^{34}SO_2}$ & 27(1,27) - 26(0,26) & 482025.14 & -2.82 & 329.28 & 63.92 & 0.23 $\pm$ 0.03 & 6.52 $\pm$ 0.95 & 1.62 $\pm$ 0.31 & 0.0004\\ 
$\rm{^{34}SO_2}$ & 7(4,4) - 6(3,3) & 479309.99 & -3.06 & 63.70 & 10.25 & 0.15 $\pm$ 0.04 & 6.28 $\pm$ 1.95 & 1.02 $\pm$ 0.42 & 0.0002\\ 
\cline{1-10}
\multicolumn{10}{|c|}{Ep Aqr}\\
\cline{1-10}
SO$_2$ & 10(3,7) - 10(0,10) & 479317.52 & -4.92 & 72.71 & 0.20 & 0.14 $\pm$ 0.02 & 10.28 $\pm$ 2.10 & 1.50 $\pm$ 0.41 & 0.0008\\ 
SO$_2$ & 19(9,11) - 20(8,12) & 481166.24 & -3.83 & 372.91 & 4.44 & 0.15 $\pm$ 0.04 & 6.95 $\pm$ 2.26 & 1.09 $\pm$ 0.47 & 0.0009\\ 
SO$_2$ & 25(2,24) - 24(1,23) & 482503.15 & -2.98 & 303.69 & 40.61 & 1.52 $\pm$ 0.06 & 8.63 $\pm$ 0.41 & 13.91 $\pm$ 0.88 & 0.0094\\ 
SO$_2$ & 7(4,4) - 6(3,3) & 491934.72 & -3.02 & 65.01 & 10.27 & 2.10 $\pm$ 0.08 & 5.55 $\pm$ 0.25 & 12.37 $\pm$ 0.75 & 0.0130\\ 
SO$_2$ & 12(3,9) - 11(2,10) & 494779.73 & -3.26 & 93.96 & 9.80 & 1.81 $\pm$ 0.07 & 6.74 $\pm$ 0.29 & 12.92 $\pm$ 0.73 & 0.0112\\
 $\rm{^{34}SO_2}$ & 27(1,27) - 26(0,26) & 482025.14 & -2.82 & 329.28 & 63.92 & 0.10 $\pm$ 0.02 & 8.02 $\pm$ 1.92 & 0.88 $\pm$ 0.38 & 0.0006\\ 
\cline{1-10}
\hline
\end{longtable}

\begin{table}[hbtp]
    \small
    \centering
    \caption{Species detected towards {\it o} Ceti arc-like structure}
    \begin{tabular}{|c|c|c|c|c|c|}
    \hline
    Species & QNs & Freq & log$_{10}(\rm{A_{ij}})$ & E$_{up}$ &  $\rm{S_{ij}\mu^{2}}$\\ 
            &     & (MHz) & (s$^{-1}$) & (K) & (D$^2$)\\
    \hline
    SO$_2$ & 4(3,1) - 3(2, 2) & 332505.24& -3.48 & 31.29 & 6.92\\ 
    SO$_2$  & 7(4,4) - 6(3,3) & 491934.72 & -3.02 & 65.01 & 10.27 \\ 
    SO$_2$  & 12(3,9) - 11(2,10) & 494779.73 & -3.26 & 93.96 & 9.80\\ 
    SO$_2$ & 16(0,16) - 15(1,15) & 283464.77 & -3.57 & 121.03 & 33.60 \\ 
    SO$_2$ & 20(1,19) - 20(0,20) & 282292.81 & -4.00 & 198.88 & 15.69\\ 
    SO & 7(6) - 6(5) & 296550.06 & -3.49 & 64.89 & 13.83 \\ 
    SO & 8(8) - 7(7) & 344310.61& -3.29 & 87.48 & 18.56 \\ 
    $^{30}$SiO&J = 7-6&296575.74&-2.86&56.94&67.18\\
    Si$^{18}$O&J = 7-6&282434.73&-2.93&54.22&67.18\\
    $^{29}$SiO&J = 8-7&342980.84&-2.67&74.08&76.79\\
    CO &J = 3-2&345795.99&-5.62&55.31&0.05\\
    $^{13}$CO&J = 3-2&330587.96&-5.95&31.73&0.07\\
    PO &J = 13/2-11/2, $\Omega$ = 1/2, F= 7- 6, l=e&283586.81&-3.91&50.26&6.92\\
    PO &J = 13/2-11/2, $\Omega$ = 1/2, F= 6- 5, l=f&283785.40&-3.91&50.32&5.92\\
    \hline
    \end{tabular}
    \label{tab:lines-oceti-arc}
\end{table}

\section{Observed and Synthetic Spectra}
Here we present the synthetic spectra obtained with MCMC fitting and the observed spectra for R Dor, R Leo, W Hya, and Ep Aqr.

\begin{figure*}[h]
    \centering
    \includegraphics[width=\textwidth]{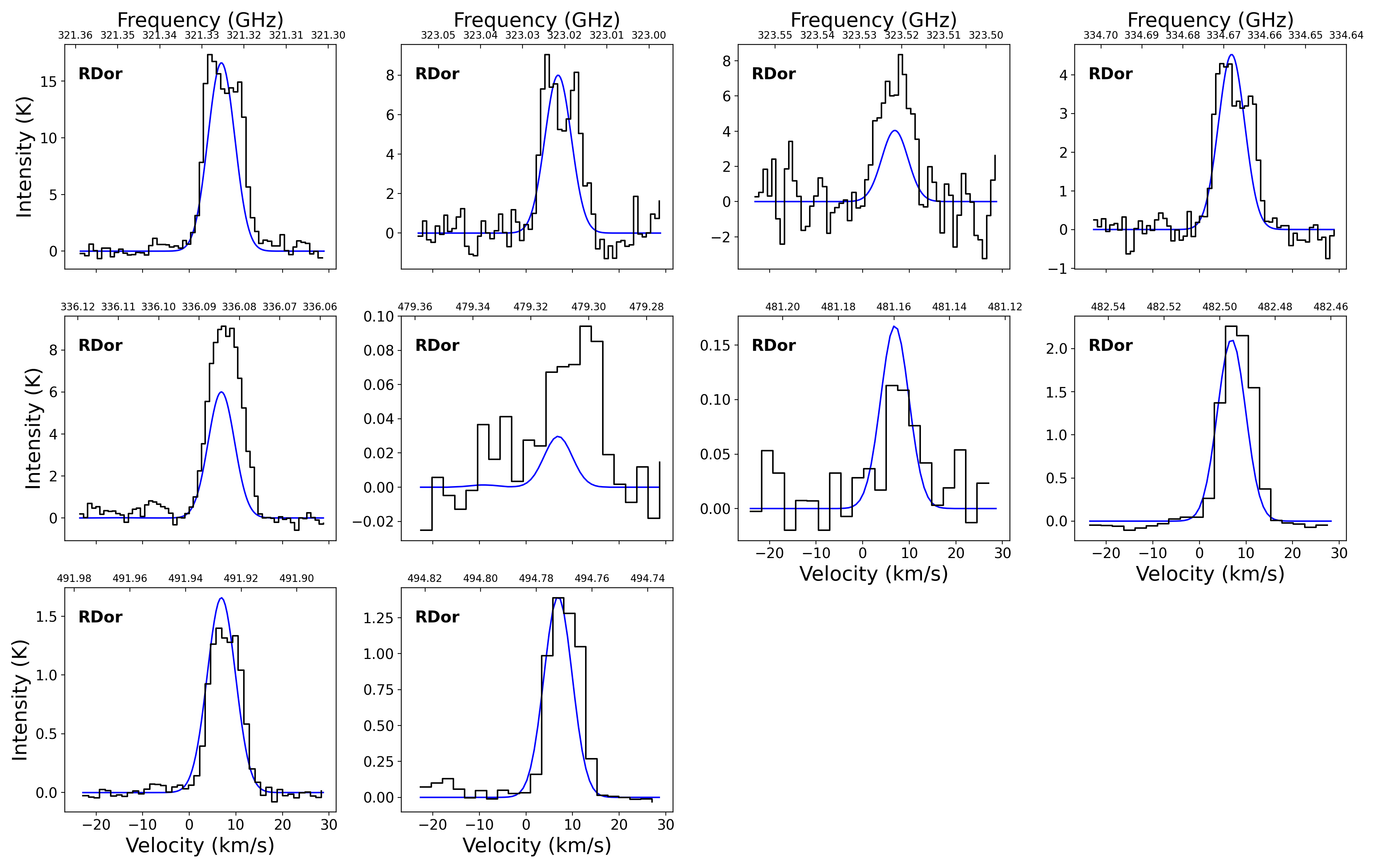}
    \caption{Observed and synthetic spectra of $\rm{SO_2}$ transitions towards R Dor. The black line represents the observed spectrum, while the blue line represents the modeled spectrum.}   
    \label{fig:synth-rdor-rleo}
\end{figure*}

\begin{figure*}[h]
    \centering
    \includegraphics[width=0.99\textwidth]{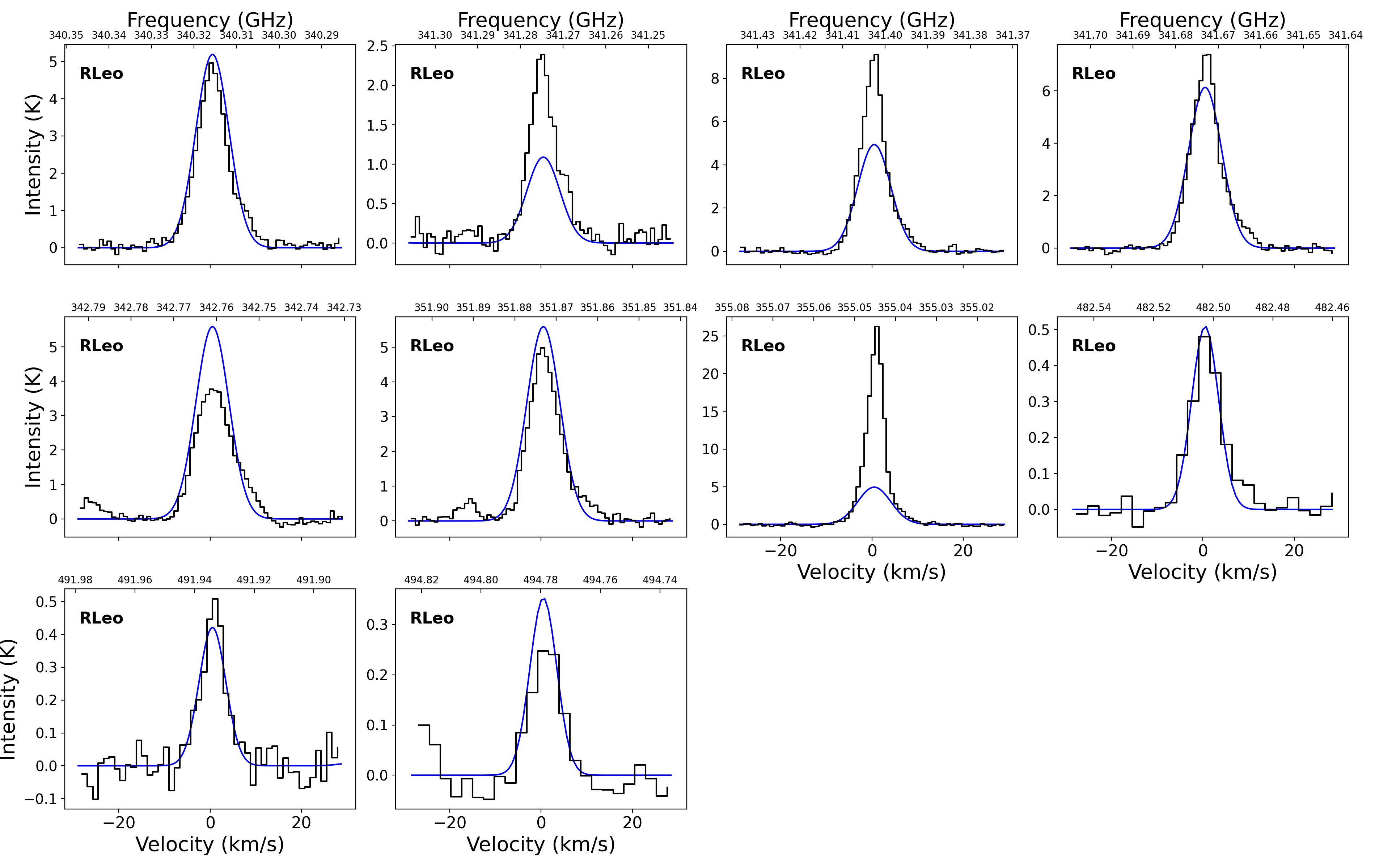}
    \includegraphics[width=\textwidth]{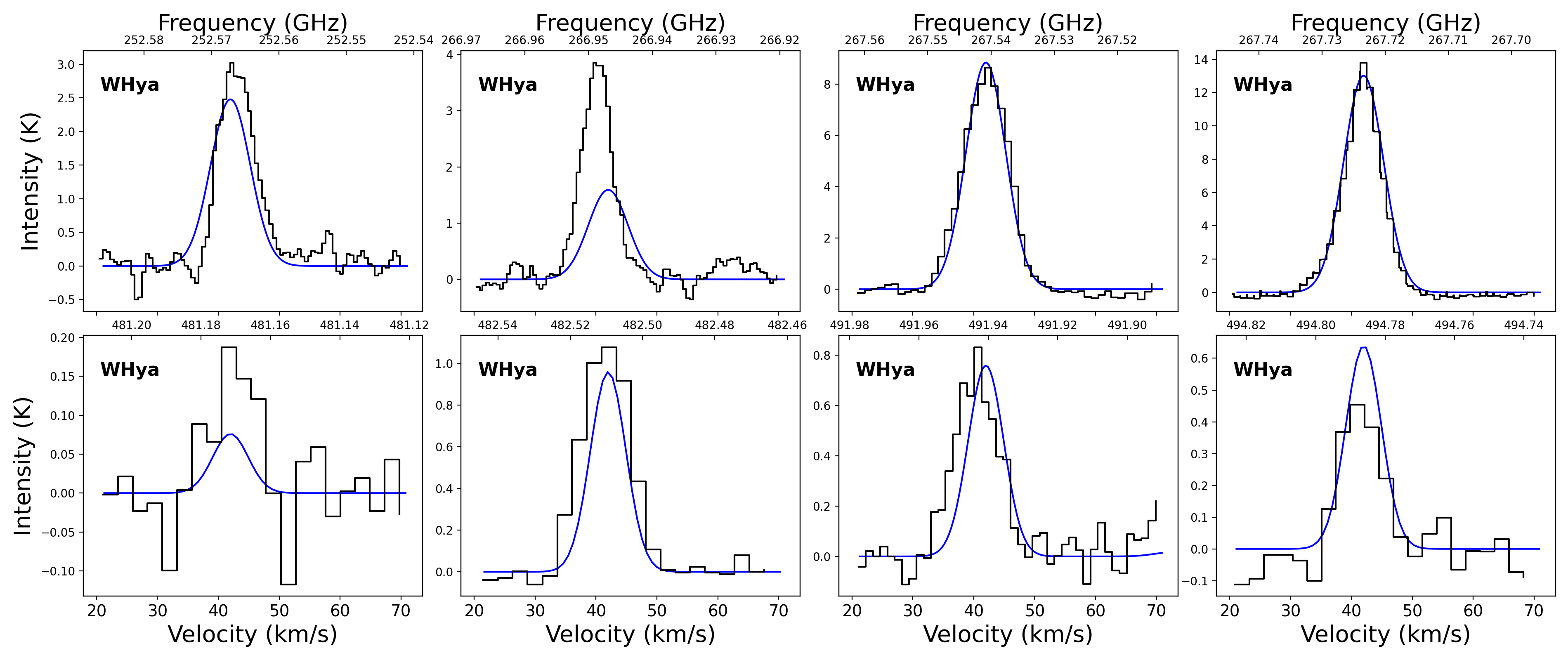}
    \includegraphics[width=\textwidth]{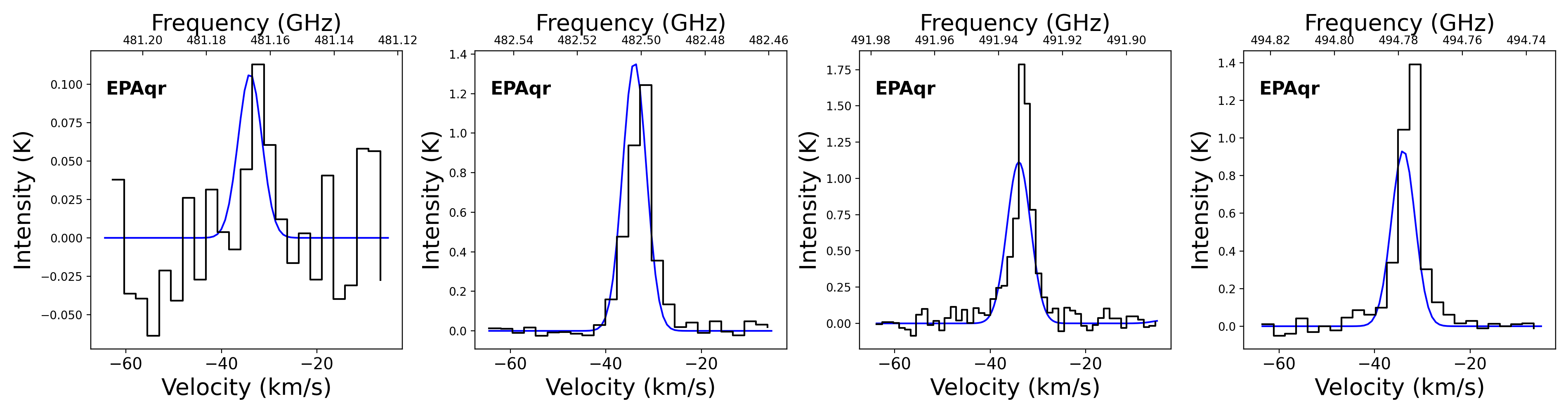}
    \caption{Observed and synthetic spectra of $\rm{SO_2}$ transitions towards R Leo, W Hya, and E PAqr. The black line represents the observed spectrum, while the blue line represents the modeled spectrum.}   
    \label{fig:synth-whya-epaqr}
\end{figure*}

\FloatBarrier
\section{Moment 0 and channel maps}
\begin{figure*}[h]
    \centering
    \includegraphics[width=\textwidth]{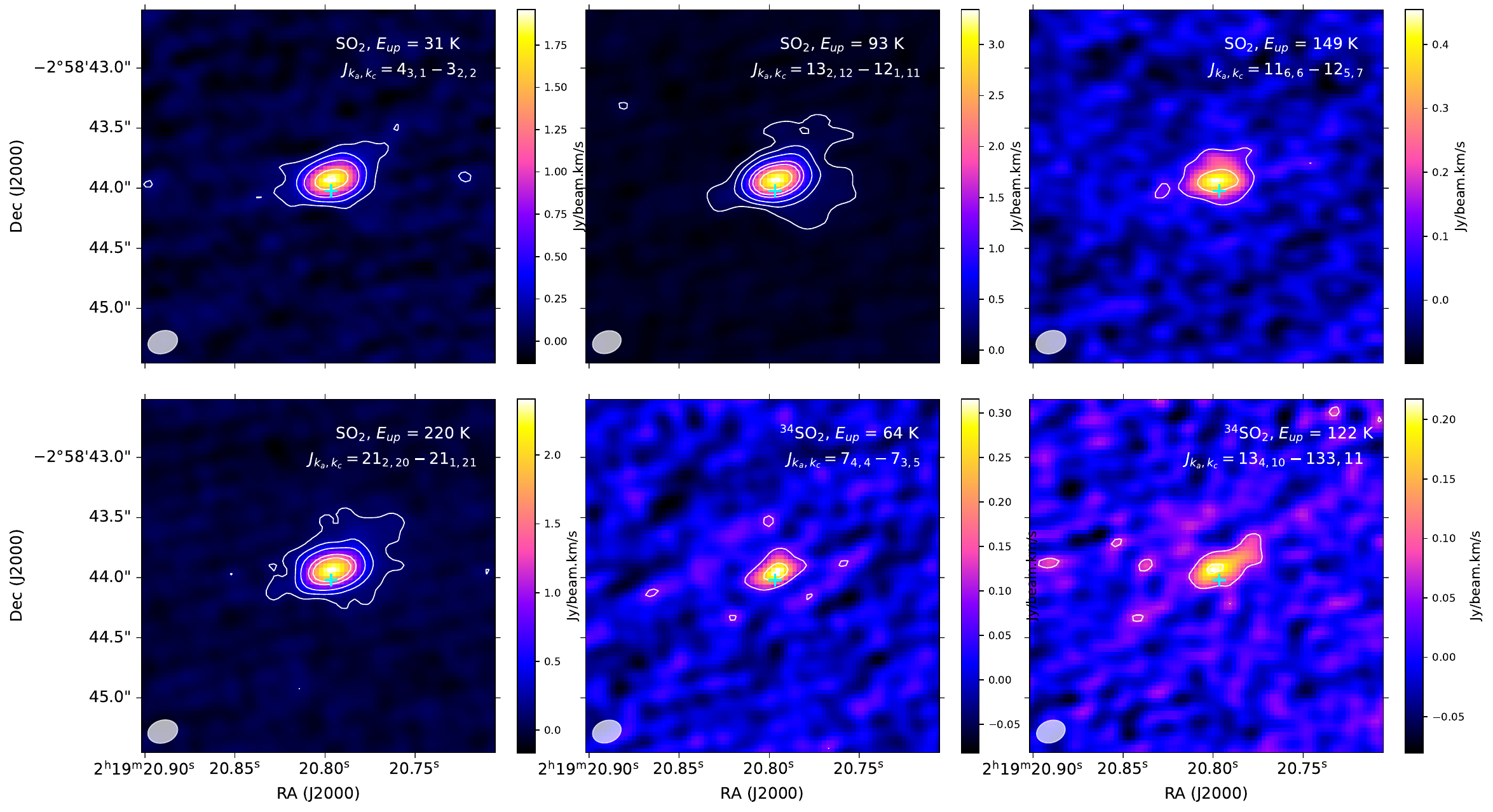}
    \caption{Moment 0 maps of SO$_2$ transitions using ALMA Band 7 observations towards source {\it o} Ceti. Contours are drawn with steps of 3$\sigma$,  9$\sigma$,  18$\sigma$,  36$\sigma$,  48$\sigma$,  72$\sigma$, and 96$\sigma$. The beam size is shown at the lower left corner of each subplot as a filled grey ellipse. The cyan cross symbol indicates the position of the continuum peak.}   
    \label{fig:mom0-octei-band7}
\end{figure*}

\begin{figure*}[h]
    \centering
    \includegraphics[width=\textwidth]{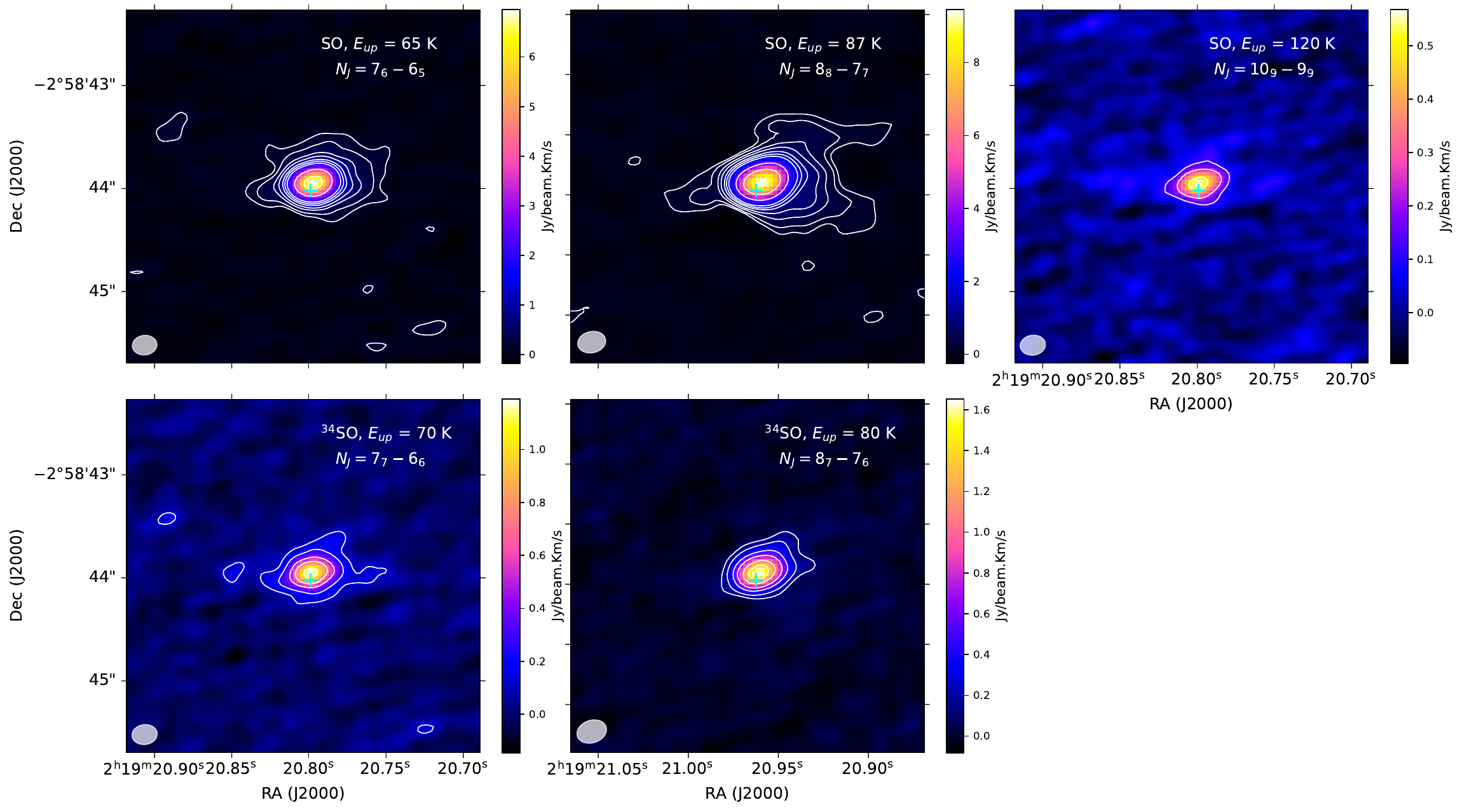}
    \caption{Moment 0 maps of SO and $\rm{^{34}SO}$ transitions using ALMA Band 7 observations towards source {\it o} Ceti. Contours are drawn with steps of 3$\sigma$,  9$\sigma$,  18$\sigma$,  36$\sigma$,  48$\sigma$,  and 72$\sigma$. The beam size is shown at the lower left corner of each subplot as a filled grey ellipse. The cyan cross symbol indicates the position of the continuum peak.}   
    \label{fig:mom0-octei-so-34so}
\end{figure*}

\begin{figure*}[h]
    \centering
    \includegraphics[width=\textwidth]{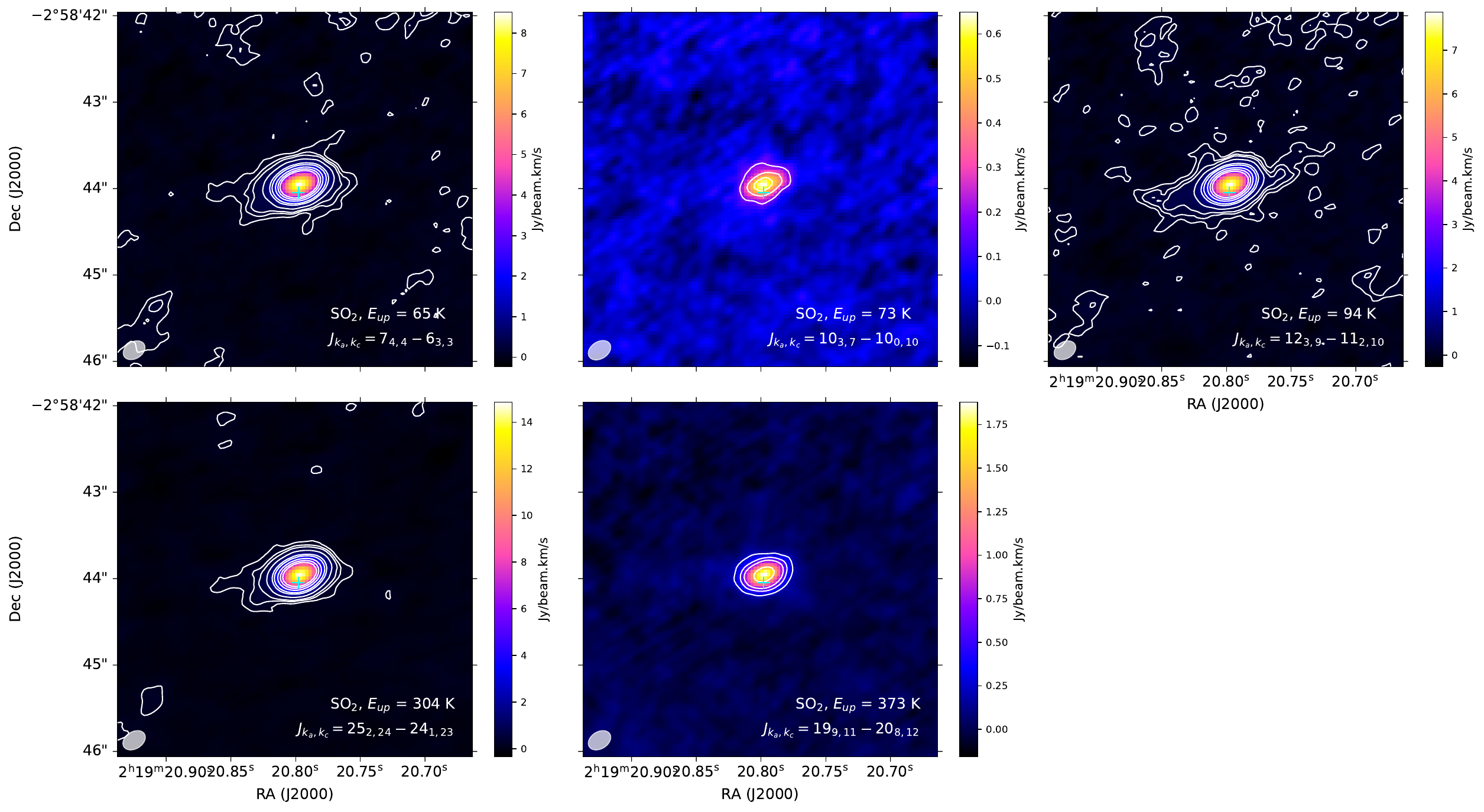}
    \caption{Moment 0 maps of SO$_2$ transitions using ALMA Band 8 observations towards source {\it o} Ceti. Contours are drawn with steps of 3$\sigma$,  9$\sigma$,  18$\sigma$,  36$\sigma$,  48$\sigma$,  72$\sigma$, and 96$\sigma$. The beam size is shown at the lower left corner of each subplot as a filled grey ellipse. The cyan cross symbol indicates the position of the continuum peak.}   
    \label{fig:mom0-oceti-band8}
\end{figure*}

\begin{figure*}[h]
    \centering
    \includegraphics[width=\textwidth]{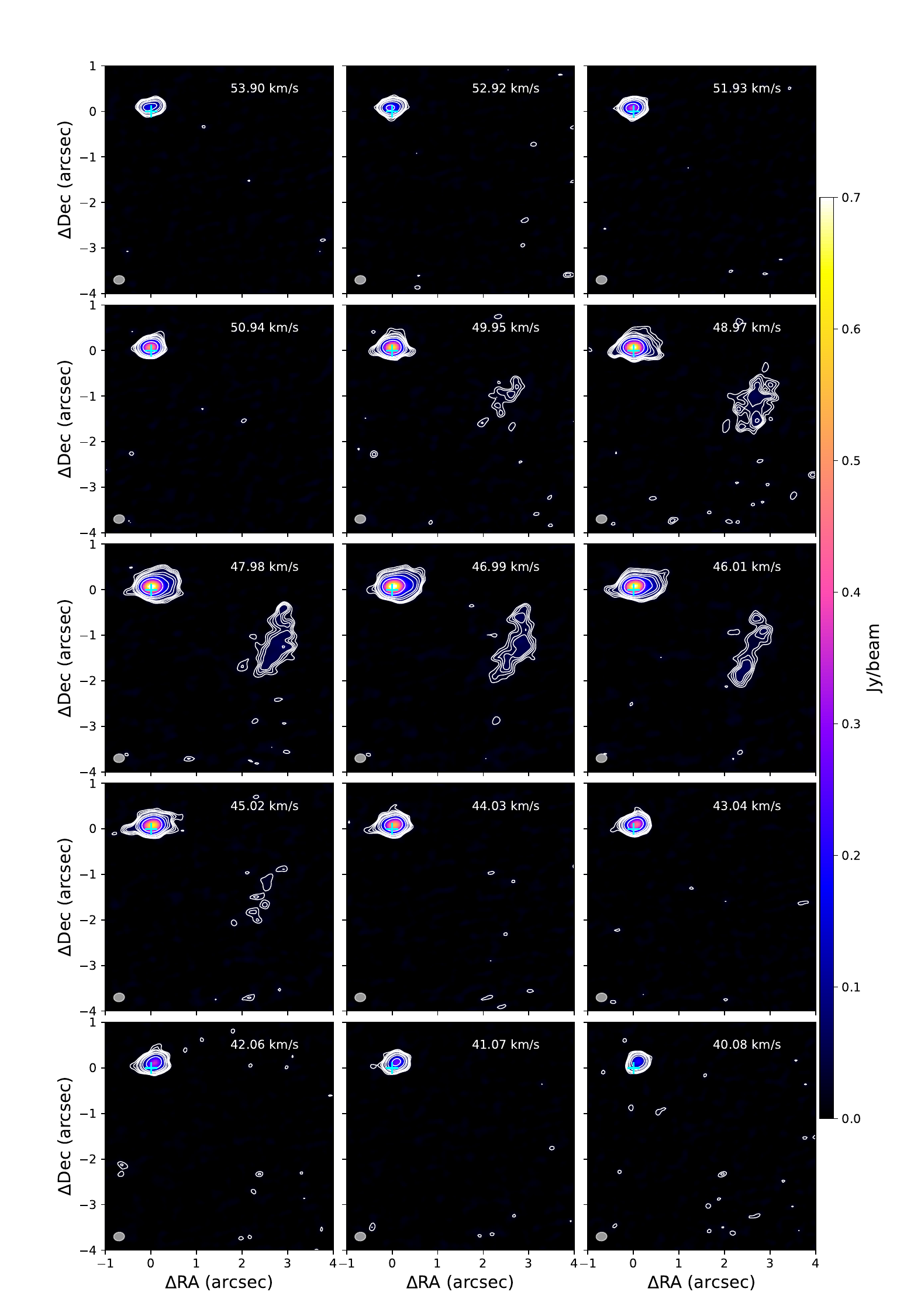}
    \vspace{-1.0cm}
    \caption{Channel maps of SO (${\rm J_{K_a,\ K_c}}$ = 6$_7$ - 5$_6$) low excitation transition (E$_{up}$ = 65 K) towards {\it o} Ceti. Contours are drawn with steps of 3$\sigma$,  4$\sigma$,  5$\sigma$,  6$\sigma$, 9$\sigma$,  12$\sigma$, 18$\sigma$, and 36$\sigma$. The beam size is shown at the lower left corner of each subplot as a filled grey ellipse. The cyan cross symbol indicates the position of the continuum peak.}  
    \label{fig:channel-maps-oceti-SO}
\end{figure*}

\begin{figure*}[h]
    \centering
    \vspace{-2.0cm}
    \includegraphics[width=\textwidth]{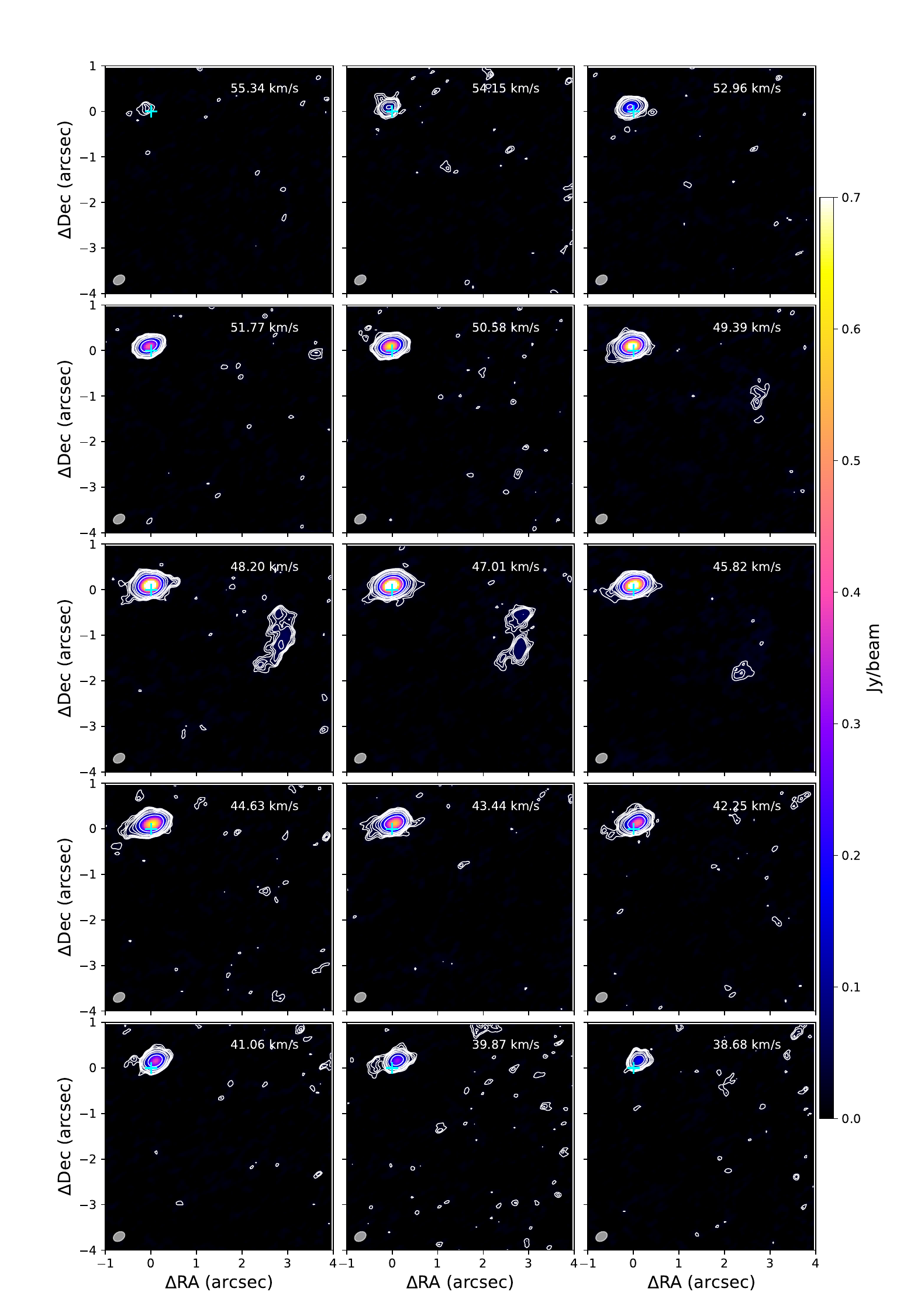}
    \vspace{-1.0cm}
    \caption{Channel maps of SO$_2$ (${\rm J_{K_a,\ K_c}}$ = 7$_{4,4}$ - 6$_{3,3}$) low excitation transition (E$_{up}$ = 65 K) towards {\it o} Ceti. Contours are drawn with steps of 3$\sigma$,  4$\sigma$,  5$\sigma$,  6$\sigma$, 9$\sigma$,  12$\sigma$, 18$\sigma$, and 36$\sigma$. The beam size is shown at the lower left corner of each subplot as a filled grey ellipse. The cyan cross symbol indicates the position of the continuum peak.}
    \label{fig:channel-maps-oceti-SO2}
\end{figure*}

In the appendix, we present moment 0 maps of all detected strong transitions, as well as channel maps of SO and SO$_2$, which are provided to illustrate the extended atmosphere of {\it o} Ceti. Figures~\ref{fig:mom0-octei-band7}, \ref{fig:mom0-octei-so-34so}, and \ref{fig:mom0-oceti-band8} show the moment 0 maps of the detected transitions toward {\it o} Ceti (one additional figure (Fig. 1) of moment 0 maps of SO$_2$ is provided in the supplementary document). The channel maps of SO and SO$_2$ are presented in Figures~\ref{fig:channel-maps-oceti-SO} and \ref{fig:channel-maps-oceti-SO2}, respectively. Moment 0 maps of the detected transitions toward R Dor are shown in Figure \ref{fig:mom0-Rdor-12m} (one additional figure (Fig. 2) of moment 0 maps of SO$_2$ is provided in the supplementary document). For R Leo, the moment 0 maps are shown in Figure \ref{fig:mom0_RLeo_12m} and \ref{fig:mom0_RLeo_so-34so-34so2} (one additional figure (Fig. 3) of moment 0 maps of SO$_2$ is provided in the supplementary document). Figure ~4 in supplementary document and Fig. \ref{fig:mom0_Whya_12m} display the moment 0 maps of the transitions identified toward W Hya. Moment 0 maps of the detected transitions toward EP Aqr are available in the supplementary document (Fig. 5).

\begin{figure*}[h]
    \centering
    \includegraphics[width=0.90\textwidth]{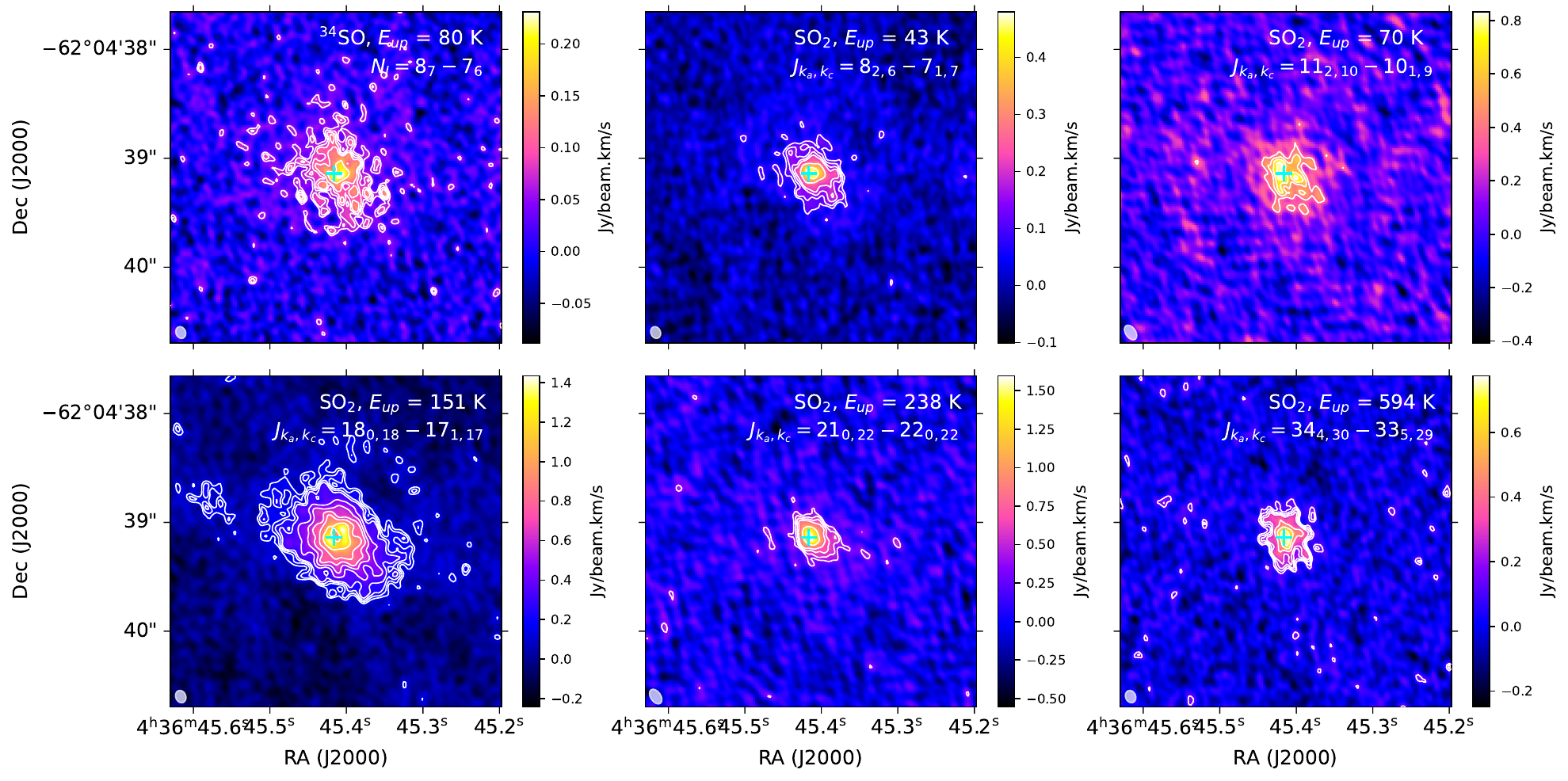}
    \caption{ Moment 0 maps of SO$_2$ transitions towards R Dor using ALMA 12m array observations. Contours are drawn with steps of 3$\sigma$,  4$\sigma$,  5$\sigma$,  6$\sigma$, 9$\sigma$,  12$\sigma$, 15$\sigma$, 18$\sigma$, and 24$\sigma$. The beam size is shown at the lower left corner of each subplot as a filled grey ellipse. The cyan cross symbol indicates the position of the continuum peak.}  
    \label{fig:mom0-Rdor-12m}
\end{figure*}

\begin{figure*}[h]
    \centering
    \includegraphics[width=0.9\textwidth]{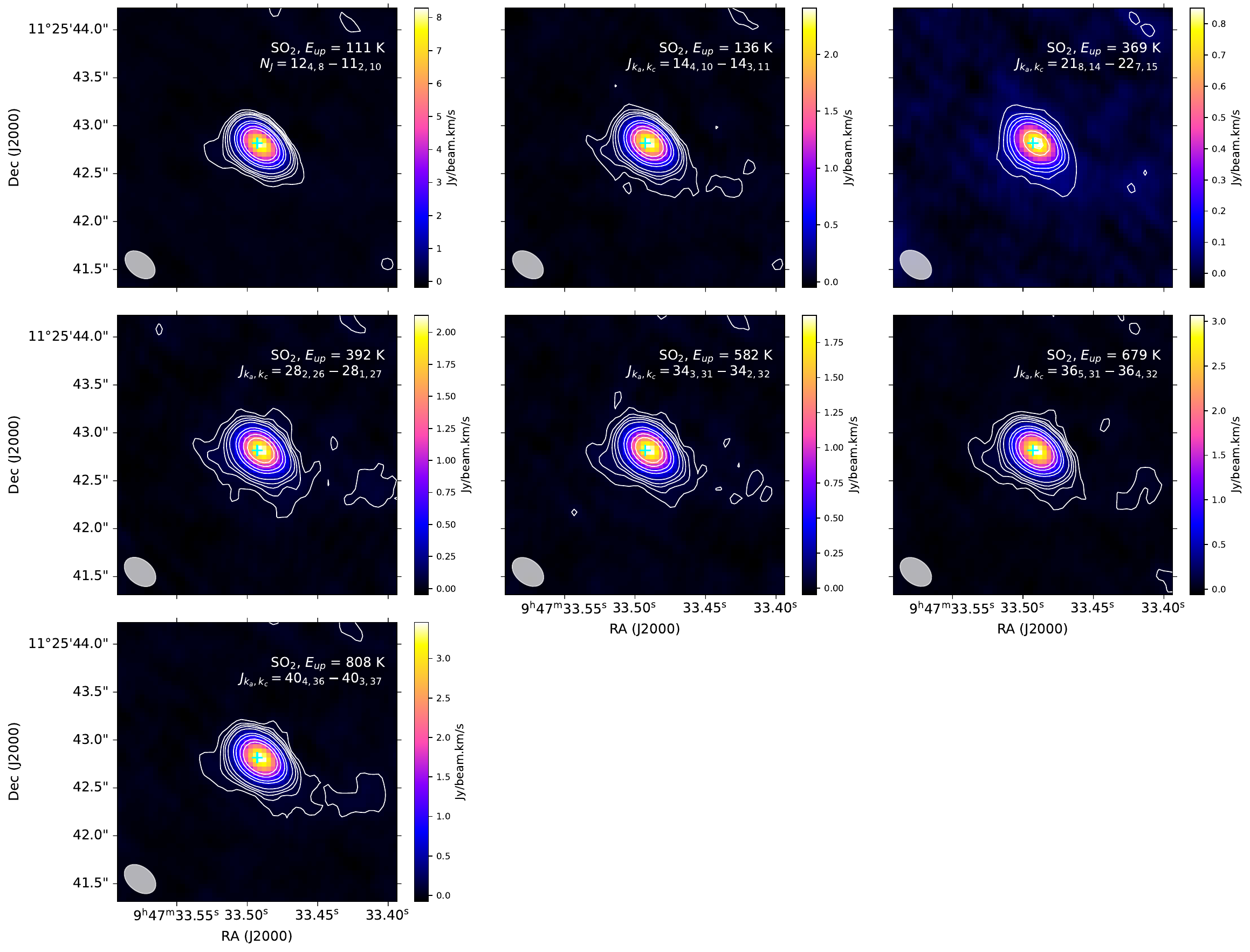}
    \caption{Moment 0 maps of SO$_2$ transitions towards R Leo using ALMA 12m array observations. Contours are drawn with steps of 3$\sigma$,  4$\sigma$,  5$\sigma$,  6$\sigma$, 9$\sigma$,  12$\sigma$, 18$\sigma$, 36$\sigma$, 48$\sigma$, 72$\sigma$, and 96$\sigma$. The beam size is shown at the lower left corner of each subplot as a filled grey ellipse. The cyan cross symbol indicates the position of the continuum peak.}   
    \label{fig:mom0_RLeo_12m}
\end{figure*}

\begin{figure*}[h]
    \centering
    \includegraphics[width=0.85\textwidth]{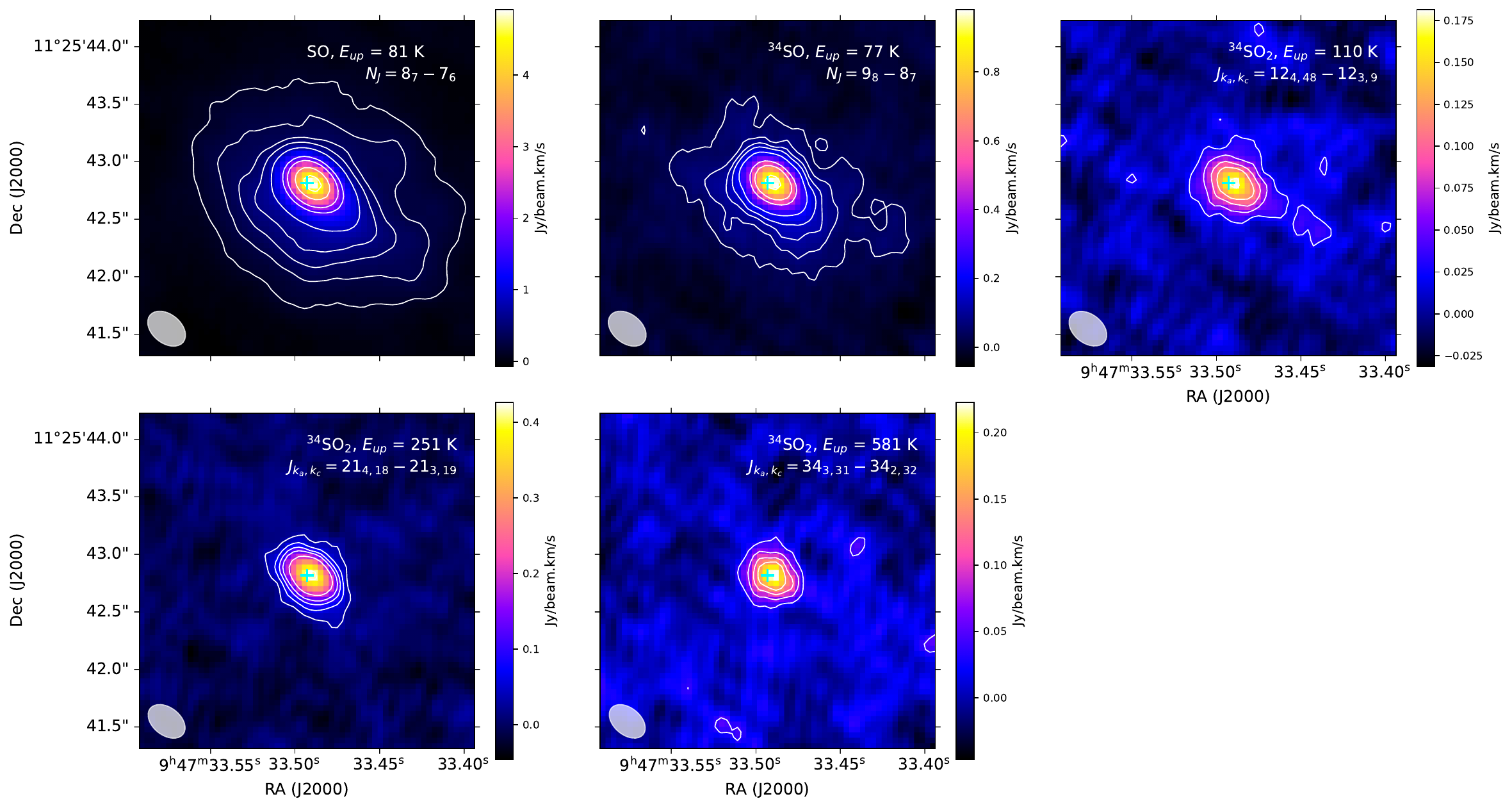}
    \caption{Moment 0 maps of SO, $\rm{^{34}SO}$, and $\rm{^{34}SO_2}$ transitions towards R Leo using ALMA 12m array observations. Contours are drawn with steps of 3$\sigma$, 6$\sigma$, 9$\sigma$,  12$\sigma$, 18$\sigma$, 36$\sigma$, 48$\sigma$, 72$\sigma$, and 96$\sigma$. The beam size is shown at the lower left corner of each subplot as a filled grey ellipse. The cyan cross symbol indicates the position of the continuum peak.}   
    \label{fig:mom0_RLeo_so-34so-34so2}
\end{figure*}

\begin{figure*}[h]
    \centering
    \includegraphics[width=0.85\textwidth]{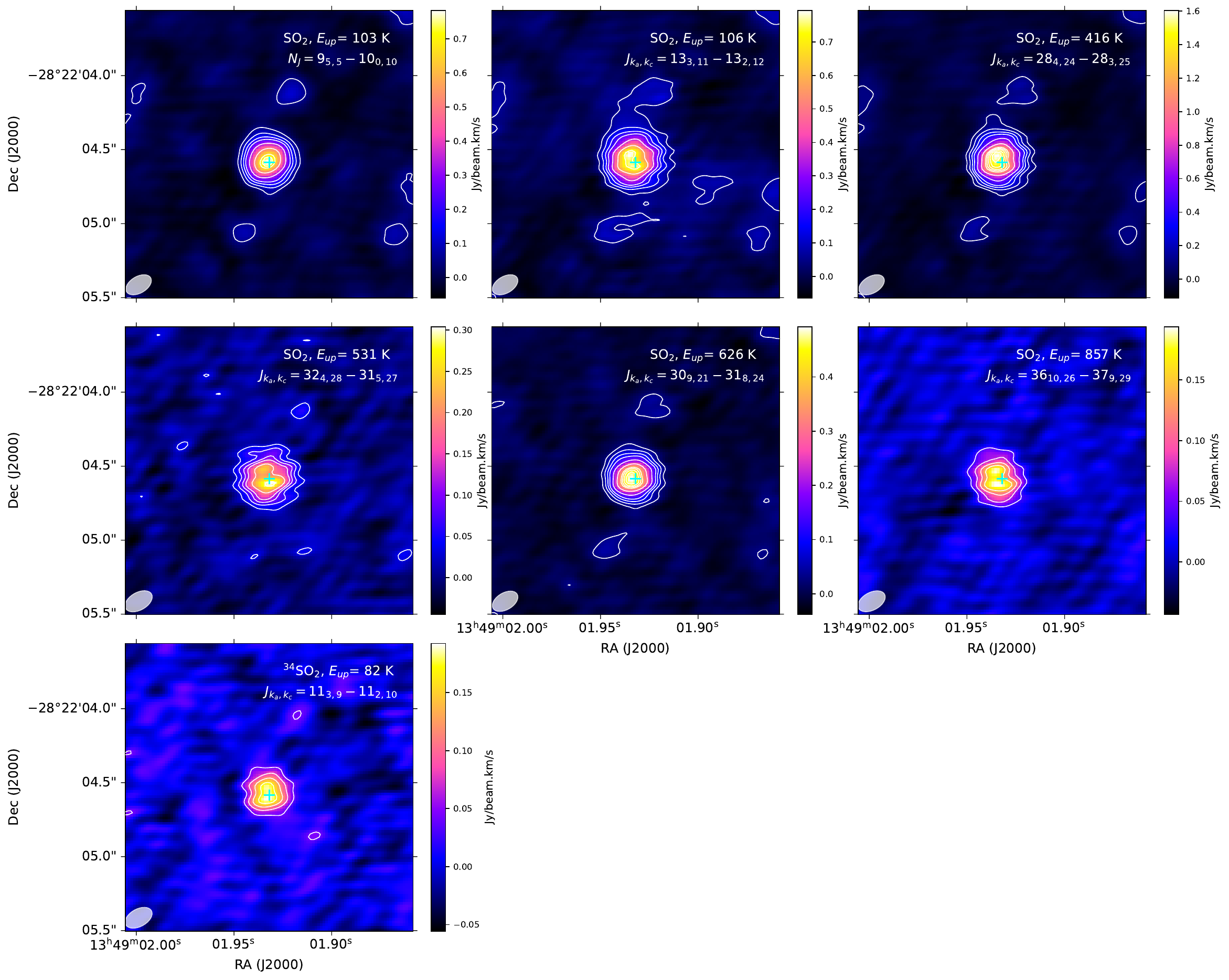}
    \caption{Moment 0 maps of SO$_2$ transitions towards W Hya using ALMA 12m array observations. Contours are drawn with steps of 3$\sigma$,  4$\sigma$,  5$\sigma$,  6$\sigma$, 9$\sigma$,  12$\sigma$, 18$\sigma$, and 36$\sigma$. The beam size is shown at the lower left corner of each subplot as a filled grey ellipse. The cyan cross symbol indicates the position of the continuum peak.}   
    \label{fig:mom0_Whya_12m}
\end{figure*}
\end{appendix}
\end{document}